\RequirePackage{fix-cm}
\documentclass[smallcondensed]{svjour3}     
\smartqed  
\usepackage{multirow}
\usepackage{array}
\usepackage[authoryear]{natbib}
\usepackage{amsmath,amssymb,amsfonts}
\usepackage{csquotes}
\usepackage{longtable}
\usepackage{caption}
\usepackage{tikz}
\usepackage{xurl}
\usepackage{float}
\newcolumntype{P}[1]{>{\raggedright\arraybackslash}p{#1}}
\usepackage[colorlinks=true,linkcolor=black,anchorcolor=black,citecolor=black,filecolor=black,menucolor=black,runcolor=black,urlcolor=black]{hyperref}

\journalname{Artificial Intelligence Review}
\begin{document}
	
	\title{Modality specific U-Net variants for biomedical image segmentation: A survey}
	
	\titlerunning{Modality specific U-Net variants for biomedical image segmentation: A survey}        
	
	\author{Narinder Singh Punn*         \and
		Sonali Agarwal 
	}
	
	\authorrunning{Punn N S, Agarwal S} 
	
	\institute{Narinder Singh Punn* \at
		IIIT Allahabad, Prayagraj, India, 211015 \\
		Tel.: +91-7018466740\\
		\email{pse2017002@iiita.ac.in}           
		\and
		Sonali Agarwal \at
		IIIT Allahabad, Prayagraj, India, 211015 \\
	}
	

	\maketitle
	
	\begin{abstract}
		With the advent of advancements in deep learning approaches, such as deep convolution neural network, residual neural network, adversarial network; U-Net architectures are most widely utilized in biomedical image segmentation to address the automation in identification and detection of the target regions or sub-regions. In recent studies, U-Net based approaches have illustrated state-of-the-art performance in different applications for the development of computer-aided diagnosis systems for early diagnosis and treatment of diseases such as brain tumor, lung cancer, alzheimer, breast cancer, etc., using various modalities. This article contributes in presenting the success of these approaches by describing the U-Net framework, followed by the comprehensive analysis of the U-Net variants by performing 1) inter-modality, and 2) intra-modality categorization to establish better insights into the associated challenges and solutions. Besides, this article also highlights the contribution of U-Net based frameworks in the ongoing pandemic, severe acute respiratory syndrome coronavirus 2 (SARS-CoV-2) also known as COVID-19. Finally, the strengths and similarities of these U-Net variants are analysed along with the challenges involved in biomedical image segmentation to uncover promising future research directions in this area.
	\end{abstract}
	
	\section{Introduction}
	Biomedical information technologies have a great importance, particularly in medi-cine to solve different problems \citep{goceri2018biomedical, kaya2017automated, goceri2016automatic, gocceri2015comparative, gocceri2013comparative}, and deep learning-based approaches have been widely used recently \citep{goceri2021diagnosis, gocceri2020impact}. The evolving medical imaging acquisition system~\citep{alexander2019scanning} has brought the consideration of the research community towards the non-invasive practice of disease diagnosis. Every diagnostic procedure involves the careful and critical examination of medical scans which represents the complex interior structure within the body, illustrating the functioning of various organs.	With a wide variety of medical imaging such as magnetic resonance imaging (MRI), X-ray, computerized tomography/computerized axial tomography (CT/ CAT), ultrasound (US), positron emission tomography (PET), etc., the medical domain has experienced exponential growth in the diagnosis practices. Each of these scans varies in the imaging procedure, usecases and its average diagnosis duration~\citep{MITM, TMI}, as shown in Table~\ref{tab1}. For any radiologist, analyzing such complex scans is tedious and time consuming, thereby to fill this void of complexity, deep learning approaches are well explored to address the automated assistance in diagnosis procedure, resulting in faster and better practices for monitor, cure and treatment of the diseases~\citep{elnakib2011medical, masood2015survey, deepa2011survey, maintz1998survey}. 
	
	Segmentation~\citep{minaee2020image} is one such automation task that helps to identify and detect the desired regions or objects of interest for the concerned issue. Depending on the depth of identifying the classes of objects, segmentation is divided into two levels as semantic and instance. The semantic segmentation~\citep{liu2019recent} segregates the objects belonging to different classes, whereas instance segmentation~\citep{chen2019instance} goes deeper to also segregate the objects within the common class. With the exhaustive analysis~\citep{minaee2020image, haque2020deep}, it is observed that among the latest advancements to perform segmentation, mostly U-Net~\citep{ronneberger2015u} based frameworks are adopted to achieve state-of-the-art segmentation performance which follows from its symmetrical encoder-decoder structure to extract and reconstruct the feature maps. 
	
	\begin{table}[]
		\centering
		\caption{Medical imaging approaches for diagnosis.}
		
		\label{tab1}
		\resizebox{0.95\columnwidth}{!}{
			\begin{tabular}{lp{0.6in}p{1.8in}l}
				\hline
				\multicolumn{1}{l}{\multirow{2}{*}{\textbf{Imaging type}}} & \multicolumn{1}{l}{\multirow{2}{*}{\textbf{Approach}}}          & \multicolumn{1}{l}{\multirow{2}{*}{\textbf{Usecase}}}                                                                   & \textbf{Duration} \\
				&            &                                                                    & \textbf{(in min.)} \\
				\hline
				MRI          & Magnetic fields and radio waves     & Multiple sclerosis, stroke,   tumors, spinal cord disorders, etc.           & 45 – 60            \\ 
				X-ray        & Ionizing radiation & Fractures, arthritis,   osteoporosis, breast cancer, etc.                   & 10 – 15            \\ 
				CT/CAT       & Ionizing radiation & Trauma injuries, tumors and   cancers, vascular and heart diseases, etc.    & 10 – 15            \\ 
				US           & Sound waves        & Gallbladder illness, breast lumps,   genital disorder, joint problems, etc. & 30 – 60            \\
				PET          & Radioactive tracer & Alzheimer, epilepsy, seizures,   parkinsons’ disease, etc.                  & 90 – 120           \\
				\hline
			\end{tabular}
		}
	\end{table}
	
	\begin{table}[!h]
		\centering
		\caption{Summary of existing review articles for biomedical image analysis.}
		\label{tab2}
		\begin{tabular}{P{0.8in}P{3.3in}} 
			\hline
			\textbf{Author}              & \textbf{Contribution}                                                                     \\ 
			\hline
			\cite{havaei2016deep}                    & Reviews CNN
			based approaches along with the key challenges for brain pathology
			segmentation using MRI such as tumor, lesion, etc.                                                     \\
			\cite{razzak2018deep}                    & Explores the potential
			of deep learning based approaches for various medical imaging applications
			across different modalities.                                                        \\ 
			\cite{hesamian2019deep}                  & Investigates
			state-of-the-art deep learning techniques for medical image segmentation along
			with the network training techniques and state-of-the-art solutions to the
			challenges.  \\ 
			\cite{taghanaki2021deep}                 & Covers the
			comprehensive analysis of deep learning approaches in the segmentation of
			natural and medical images with different categories and applications.                           \\ 
			\cite{zhou2019review}                    & Explores
			multi-modality fusion based approaches for medical image segmentation.  
			\\ 
			\cite{chen2020deep}                      & Reviews deep
			learning approaches for cardiac image segmentation using ultrasound, CT and
			MRI imaging along with the various techniques to address the challenges.                     \\ 
			\cite{haque2020deep}                     & Presents literature
			survey of deep learning technologies for biomedical image segmentation with
			different modalities.                                                                 \\ 
			\cite{lei2020medical}                    & Reviews
			various deep learning models for medical image segmentation with supervised
			and weakly supervised learning aspects.                                                           \\ 
			\cite{renard2020variability}             & Emphasizes the
			variability and reproducibility of the deep learning approaches for medical
			image segmentation.                                                                        \\
			\hline
		\end{tabular}
	\end{table}
	
	\subsection{Motivation and contribution}
	With recent developments in deep learning technologies, there are a lot of review articles on biomedical image segmentation (BIS) using deep learning. The understanding of the available methods is critical for developing computer-aided diagnosis systems; however, to contribute to this domain as a researcher, one needs to understand the underlying mechanics of the methods that make those systems achieve promising results. For instance, the work of \cite{zhou2019review} explored the comprehensive analysis focused on multi-modality fusion approaches, whereas \cite{haque2020deep} reviewed the standard deep learning approaches for BIS using different modalities. Table~\ref{tab2} shows the overview of some of the survey articles proposed for biomedical image analysis using deep learning approaches. In contrast to the existing review articles, the present article is intended to contribute for an exhaustive analysis of the state-of-the-art modality specific U-Net based approaches by performing inter-modality and intra-modality categorization, and establishing better insights of technological solutions for each modality. Furthermore, the present article also uncovers general and modality specific challenges to perform biomedical image segmentation and make the researchers or readers reap the most benefits from the current advancements in U-Net and aid in further contributions towards the research in this area.
	
	\subsection{Review process}
	The basis of including a research article in this survey is that the article describes the research on U-Net based biomedical image segmentation. The articles confirming vivid architectures or frameworks are only included if the authors claimed certain advancements or novel contributions, whereas articles with pure discussions are excluded; fortunately, such articles are limited and hence will not affect the outcome of this survey. The search for the articles is performed on Google Scholar, which is one of the best academic search engine~\citep{academicse2020}, where relevant articles are identified using the search string, SS1 as shown in Table~\ref{tab3}. Among the acquired papers, the high quality journals or conferences are confirmed by analyzing its impact factor (high), h-index (high), peer-review process (transparent), indexing (MEDLINE, Elsevier Scopus and EMBASE, Clarivate Analytics Web of Science, Science Citation Index, etc.) and scientific rigor. These reputed journals are identified from the ranked list, CORE~\citep{core2020}. However, some articles are also included from popular preprint servers such as arXiv. With such a huge pool of acquired articles, the most relevant articles are filtered with a thorough examination (journal or conference quality, cite score and contribution) to include in this survey. These articles are analysed by categorizing it in one of the proposed classes and highlighting the architectural design of U-Net variant along with the achieved results and further possible improvements to address modality specific segmentation challenges. 
	
	\begin{table}
		\centering
		\caption{Search strings to acquire research papers and analyse research trend using GoogleScholar.}
		
		\label{tab3}
		\resizebox{0.95\textwidth}{!}{\begin{tabular}{lP{1.7in}lll}
				\hline
				\textbf{No.} & \textbf{Search string} & \textbf{Queried date} & \textbf{Year} & \textbf{No. of papers} \\
				\hline
				SS1 & (U-Net segmentation CT OR X-ray OR PET OR US OR MRI) & November 10, 2021 & 2015-21 & 32,530 \\
				SS2 & (biomedical image segmentation) & November 10, 2021 & 2015-21 & 172,190 \\
				SS3 & (biomedical image segmentation "U-Net") & November 10, 2021 & 2015-21 & 38,900 \\
				\hline
		\end{tabular}}
	\end{table}
	
	\begin{figure}
		\centering
		\includegraphics[width=\linewidth] {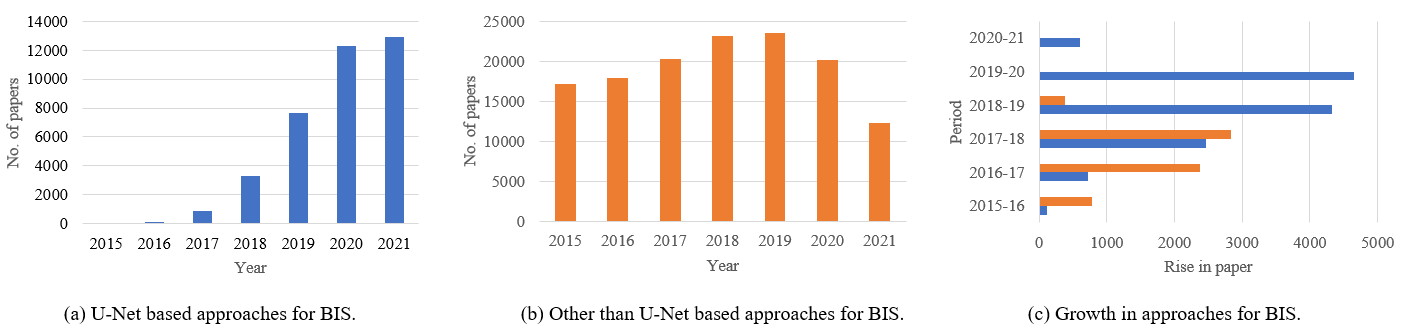}
		\caption{Research trend in biomedical image segmentation per year.}
		\label{fig1}
	\end{figure}
	
	\subsection{Research trend in BIS}
	A comparative literature exploration is conducted on the Google Scholar search engine using the search strings, SS2 and SS3, as shown in Table~\ref{tab3}. The number of BIS approaches without U-Net are acquired by subtracting the number of BIS U-Net articles from the pool of BIS articles, to understand the latest trend of research. Fig.~\ref{fig1} illustrates that the latest approaches are developed by employing the U-Net framework while experiencing exponential growth every year. In order to analyse such trend, this article aims to provide an exhaustive review of the variants of U-Net architectural design developed for segmentation. It is evident that the U-Net model incorporates the huge potential for further advancements due to its mutable and modular structure that would result in a state-of-the-art diagnosis system.
	
	\subsection{Article structure}
	The remaining portion of the article is divided into several sections, where Section \ref{sec2} presents the overview of biomedical image analysis and in Sections \ref{sec3}, \ref{sec4} and \ref{sec5} the comprehensive analysis of U-Net variants is presented that covers implementation strategies and advancements. Later, Section \ref{sec6} presents the observations concerned with the current advancements in U-Net based approaches, followed by the scope and challenges in Section \ref{sec7} and concluding remarks in Section \ref{sec8}. 
	
	\section{Biomedical image analysis}
	\label{sec2}
	The success of deep learning in image analysis has encouraged biomedical imaging researchers to investigate its potential in analyzing various medical modalities to aid clinicians in faster diagnosis and treatment of diseases or infections like the ongoing pandemic of SARS-CoV-2 (COVID-19). Following the deep learning usecases, the implication of classification can ascertain the presence or absence of disease in some modality e.g. the ground glass opacification (GGO) in the lungs via CT imaging. Furthermore, in localization, normal anatomy can be identified e.g. lungs in the CT or X-ray imaging, and later segmentation can generate refined boundaries around the GGOs to understand its impact on the anatomical structures for further analysis. Since, segmentation is an extension to classification, localization or detection, it offers very rich information about the disease and infected regions. With this interest, many architectures have been proposed for the segmentation of the targeted regions from vivid modalities~\citep{haque2020deep}. In addition, segmentation is the most widely researched application of deep learning in biomedical image analysis~\citep{litjens2017survey}, where U-Net based segmentation architectures have gained significant popularity to develop computer-aided diagnosis (CAD) systems.
	
	\subsection{Rise of segmentation architectures}
\begin{figure}
		\centering
		\includegraphics[width=\columnwidth] {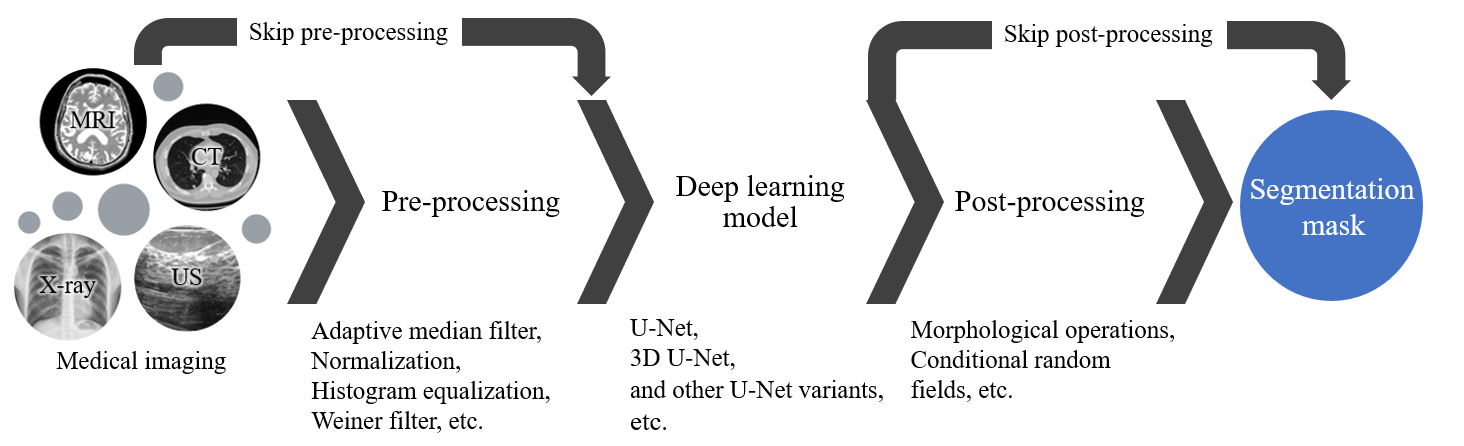}
		\caption{Schematic representation of deep learning based segmentation architectures.}
		\label{fig2}
	\end{figure}

	Despite the advancements in deep learning, segmentation is still one of the challenging tasks due to the varying dimensions, shape and locale of the target tissues. Traditionally, the segmentation process was carried manually by expert clinicians to illuminate the regions of interest in the whole volume of samples, thereby it is ideal to automate this process for faster diagnosis and treatment. In recent years, various deep learning models are developed for BIS that are categorized into manual, semi-automatic and fully automatic approaches~\citep{haque2020deep}. Fig.~\ref{fig2} presents the schematic representation of the pipeline of the recent deep learning based segmentation frameworks for biomedical images, which is divided into data preprocessing~\citep{bhattacharyya2011brief}, deep learning model~\citep{minaee2020image}, and post-processing~\citep{zhou2019tonguenet, christ2016automatic}. In the data preprocessing stage, the data undergoes a certain set of operations like resize and normalization to reduce the intensity variation in the image samples, augmentation to generate more training samples for avoiding the class biasness and overfitting problem, removal of irrelevant artefacts or noise from the data samples, etc. The pre-processed data is then fed to train the deep neural segmentation network, where mostly U-Net based architectures are deployed. The output of the network undergoes post-processing with techniques such as morphological and conditional random field based feature extraction to refine the final segmentation results. 
	
	Initiated from the sliding window approach by \cite{ciresan2012deep} to classify each pixel while also localizing the regions using patch based input, the model outperformed in the ISBI 2012 challenge; however, the training was slow because of a large number of overlapping patches and also lacked the balance of context and localization accuracy. \cite{long2015fully} proposed fully convolutional neural network (FCN) for semantic segmentation, defined on the state-of-the-art classification networks like Alex-Net, VGG-Net and Google-Net. This model can process images of arbitrary size and produce the segmentation mask of same size by using deconvolution; however, it does not utilize global information context and hence generates fuzzy segmentation masks. Later, the U-Net model proposed by \cite{ronneberger2015u}, consists of FCN along with the contraction-expansion paths and skip connections to gradually adapt the long-range affinities. The contraction phase tends to extract high and low level features, whereas the expansion phase follows from the features learned in the corresponding contraction phase (skip connections) to reconstruct the image into the desired dimensions with the help of transposed convolutions or upsampling operations. The U-Net model won the ISBI 2015 challenge and outperformed its predecessors. Later, a similar approach is proposed by \cite{cciccek20163d} in the three dimensional feature space to perform volumetric segmentation of Xenopus kidney and achieved promising results. Following from the state-of-the-art potential of the U-Net model, many variants have been proposed based on the variation in the convolution and pooling operations, skip connections, the arrangement of the components in each layer and hybrid approaches that make use of the state-of-the-art deep learning models, to tackle the challenges associated with different applications.
	
	\subsubsection{U-Net}
\begin{figure}
		\centering
		\includegraphics[scale=0.4] {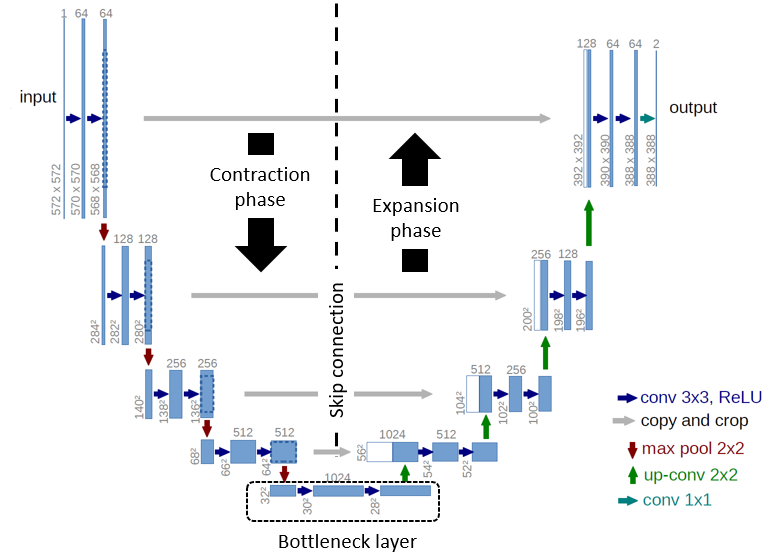}
		\caption{U-Net architecture.}
		\label{fig3}
	\end{figure}

	With the sense of segmentation being a classification task where every pixel is classified as being part of the target region or background, \cite{ronneberger2015u} proposed a U-Net model to distinguish every pixel, where input is encoded and decoded to produce output with the same resolution as input. As shown in Fig.~\ref{fig3}, the symmetrical arrangement of encoder-decoder blocks efficiently extracts and concatenates multi-scale feature maps, where encoded features are propagated to decoder blocks via skip connections and a bottleneck layer.
	
	\begin{figure}[!b]
		\centering
		\includegraphics[width=\columnwidth] {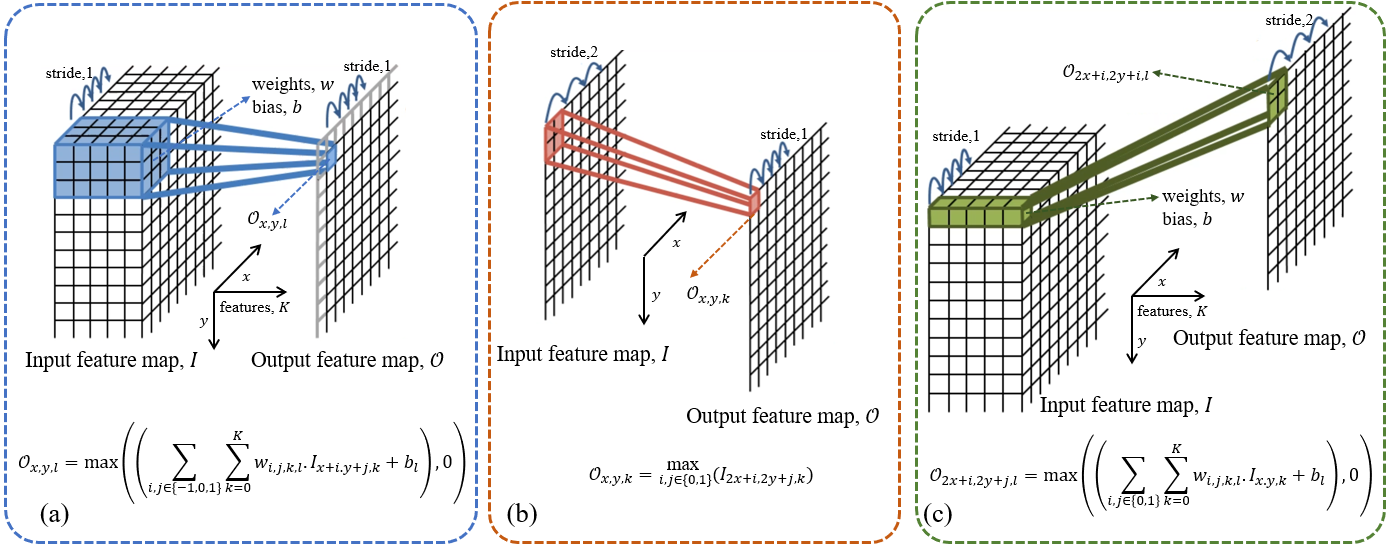}
		\caption{Summary of operations in U-Net. (a) $3\times 3$ convolution + ReLU, (b) $2\times 2$ max-pooling and (c) $2\times 2$ up-convolution operation.}
		\label{fig4}
	\end{figure}
	The encoder block (contraction path) consists of a series of operations involving valid $3\times 3$ convolution followed by a ReLU activation function (as shown in Fig.~\ref{fig4}(a)), where a 1-pixel border is lost to enable processing of the large images in individual tiles. The obtained feature maps from the combination of convolution and ReLU are downsampled with the help of max pooling operation, as illustrated in Fig.~\ref{fig4}(b). Later, the number of feature channels are increased by a factor of 2, following each layer of convolution, activation and max pooling, while resulting in spatial contraction of the feature maps. The extracted feature maps are propagated to decoder block via bottleneck layer that uses cascaded convolution layers. The decoder block (expansion path) consists of sequences of up-convolutions (as shown in Fig.~\ref{fig4}(c)) and concatenation with high-resolution features from the corresponding encoded layer. The up-convolution operation uses the kernel to map each feature vector to the $2\times 2$ pixel output window followed by a ReLU activation function. Finally, the output layer generates a segmentation mask with two channels comprising background and foreground classes. In addition, the authors addressed the challenge to segregate the touching or overlapping regions by inserting the background pixels between the objects and assigning an individual loss weight to each pixel. This energy function is represented as a pixel-wise weighted cross entropy function as shown in Eq.~\ref{eq9}. The authors established the state-of-the-art results by winning the ISBI 2015 challenge. 
	
	\begin{equation}
		E = \sum_{{x} \in \Omega}\left(w_c({x}) + w_0 \cdot \exp\left( - \frac{(d_1({x}) + d_2({x}))^2}{2\sigma^2}\right)\right) \log({p}_{\ell({x})}({x}))
		\label{eq9}
	\end{equation}
	where softmax, ${p}_k({x}) = \exp({a_k({x})}) / \left(\sum_{k' = 1}^K \exp(a_{k'}({x}))\right)$ with activation, $a_k({x})$ for channel $k$ and pixel $x\in \Omega$ with $\Omega \in \mathbb{Z}^{2}$, $w_c$ is the weight map, $d_1$ and $d_2$ are the distances to the nearest and the second nearest boundary pixels, and $w_o$ and $\sigma$ are constants.
	
	\subsubsection{Other than U-Net}
	U-Net is the most suitable segmentation model in the area of biomedical image analysis because of its ability to simultaneously combine high and low level information which helps to extract complex features and improve accuracy, respectively. However, there are various other deep learning based models that are utilized for segmentation such as FCN \citep{long2015fully}, DeepLab \citep{chen2017deeplab}, SegNet \citep{badrinarayanan2017segnet}, mask R-CNN \citep{he2017mask}, etc.
	
	\cite{long2015fully} introduced FCN that has set the foundation of segmentation architectures across various domains. In contrast to classical CNN models (VGG, ResNet, etc.) where fully connected layers are employed to categories an entire image, FCN uses 1$\times$1 convolution layers to perform pixel level classification and generate segmentation mask by upsampling the feature maps of the last convolution layer via deconvolution layer. However, with this arrangement of operations the generated masks are relatively fuzzy and insensitive to the global context information \citep{minaee2020image}. Unlike FCN which uses deconvolution to upsample the feature maps, SegNet \citep{badrinarayanan2017segnet} is designed as a symmetric encoder-decoder structure, where encoder block uses VGG16 network topology for feature extraction and a corresponding decoder block uses max pooling indices that are transferred from encoder to decoder blocks, to generate sparse upsampled feature map without using any training parameters. However, this arrangement of operations ignores the pixel adjacent information especially during upsampling of low dimensional feature maps. U-Net addresses this issue by transferring the entire feature map from encoder to decoder during upsampling, but at the cost of more memory requirement; however, it can be neglected due to the significant improvements in the segmentation results.
	
	Inspired from the potential of faster R-CNN model~\citep{ren2015faster} to perform object detection, \cite{he2017mask} proposed mask R-CNN model to further refine the object boundaries for segmentation by first computing the object detection with bounding boxes, predicting the associated classes and finally computing the binary mask to segment objects. \cite{vuola2019mask} analysed mask R-CNN model for nuclei segmentation, where the network accurately detected nuclei with bounding boxes but struggles to generate a better segmentation mask. Following this, the authors integrated mask R-CNN with U-Net to improve the overall segmentation performance.
	
	DeepLab is another family of segmentation models that have improved over the years, where each phase of enhancement is named as DeepLabv1~\citep{chen2014semantic}, DeepLabv2~\citep{chen2017deeplab}, DeepLabv3~\citep{chen2017rethinking} and DeepLabv3+~\citep{chen2018encoder}. DeepLabv1 model uses VGG16 model, where fully connected layers are removed and pooling layers are replaced with atrous convolution. DeepLabv2 model address the difficulty of the DeepLabv1 model to segment the same objects with different sizes in an image by using ResNet101 as the backbone model and atrous spatial pyramid pooling (ASPP) to capture the multi-scale context of the objects in an image. To further refine the results, in DeepLabv3 parallel or cascaded atrous convolution block is designed with multiple dilation rates to better capture multi-scale context. DeepLabv3+ further extends the DeepLabv3 with a decoder block to improve the segmentation results. It uses feature maps from the middle layer and the Xception model for segmentation. Moreover, it also uses depthwise separable convolutions with ASPP to reduce the training parameters. In most of the U-Net variants, these modules are integrated with the network to achieve better segmentation results.
	
	\begin{figure}[!b]
		\centering
		\includegraphics[width=\columnwidth] {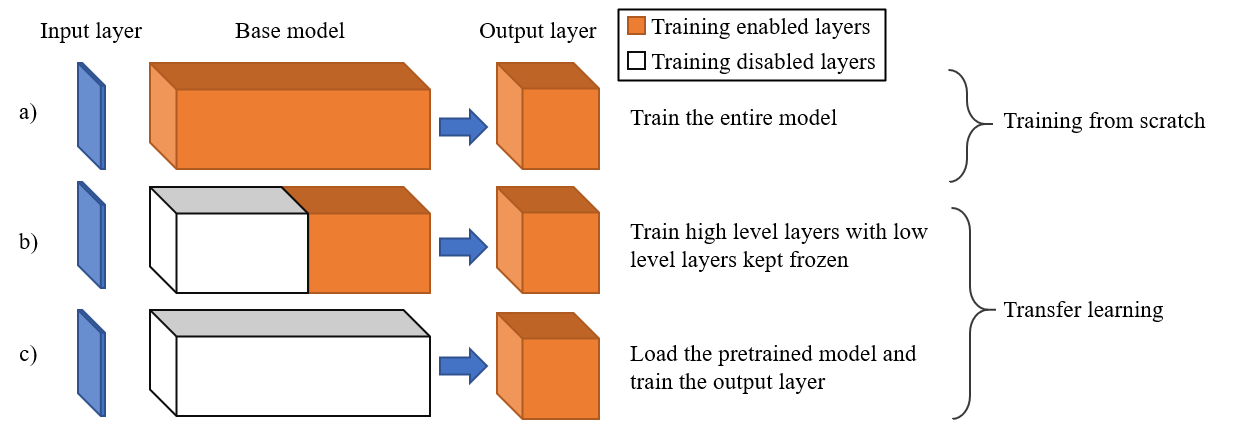}
		\caption{Typical approaches for model training.}
		\label{fig5}
	\end{figure}
	
	\subsubsection{Implementation strategies}
\begin{figure}[!t]
		\centering
		\includegraphics[scale=0.5] {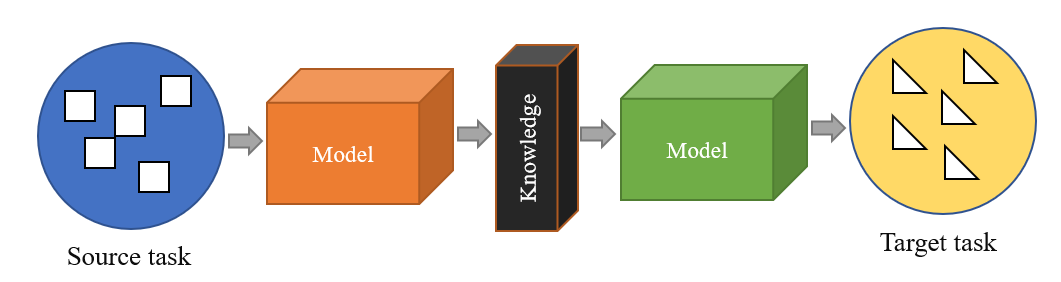}
		\caption{Illustration of transfer learning approach to adapt to new task.}
		\label{fig6}
	\end{figure}

	The implementation strategies of segmentation architectures can be divided into two categories: 1) training from scratch and 2) training using a pre-trained model (also known as transfer learning). In first approach (as shown in Fig.~\ref{fig5}(a)), an entire model is trained in which training parameters are initialized with Xavier initialization~\citep{glorot2010understanding} or Kaming initialization~\citep{he2015delving}. Due to which this approach requires a large number of labelled data samples to optimize the training parameters and learn the desired task. Hence, this approach requires intensive time and effort to develop and train the model. In the transfer learning paradigm, as simulated in Fig.~\ref{fig6}, a pre-trained model (models trained on benchmark datasets such as ImageNet) is utilized as a backbone model to train on different data involving similar or different tasks such as object detection and image segmentation. As shown in Fig.~\ref{fig5}(b) and Fig.~\ref{fig5}(c), the transfer learning or domain adaptation can be applied in two schemes, either freezing the base model and training the later layers for prediction, or semi-freezing the base model, where few high level layers are retrained along with the prediction layers. The transfer learning approach typically produces better results than the random initialization of the training parameters~\citep{garcia2018survey}.
	
	\begin{table}[!b]
		\centering
		\caption{Summary of performance metrics for BIS in terms of number of true positive ($TP$), true negative ($TN$), false positive ($FP$) and false negative ($FN$), predicted mask ($\mathcal{P}$) and ground truth ($\mathcal{G}$), $\mathcal{H}(X,Y)$ is the directed $AHD$ from $X$ to $Y$ with $d$ as euclidean distance, $\mathcal{V}_p$ and $\mathcal{V}_g$ refer to the volumes of generated and reference segmentation.}
		\label{tab4}
		\resizebox{\textwidth}{!}{\begin{tabular}{ll}
				\hline
				\textbf{Metric} & \textbf{Expression}\\
				\hline
				Accuracy & $A=\frac{(TP+TN)}{(TP+TN+FP+FN)}$\\
				Precision & $P=\frac{TP}{(TP+FP)}$\\
				Recall & $R=\frac{(TP)}{(TP+FN)}$\\ 
				F1-score & $F1=2\times\frac{(P \times R)}{(P + R)}$\\ 
				Specificity & $S=\frac{TN}{(TN+FP)}$\\ 
				
				Dice similarity coefficient & $DSC=\frac{2 \times |\mathcal{P} \cap \mathcal{G}|}{|\mathcal{P}|+|\mathcal{G}|}=\frac{2TP}{2TP+FP+FN}$\\ 
				Intersection-over-union & $IoU=\frac{\mathcal{P} \cap \mathcal{G}}{\mathcal{P}\cup \mathcal{G}}=\frac{TP}{TP+FP+FN}$\\ 
				Average Hausdo-rff distance & $AHD=\frac{1}{2}\left(\frac{\mathcal{H}(\mathcal{P},\mathcal{G})}{\mathcal{P}}+\frac{\mathcal{H}(\mathcal{G},\mathcal{P})}{\mathcal{G}}\right)$\\
				&\;\;\;\;\;\;\;\;\;\;$=\frac{1}{2}\left(\frac{1}{\mathcal{P}} \sum_{p\in \mathcal{P}} \min_{g\in \mathcal{G}} d(p,g) + \frac{1}{\mathcal{G}} \sum_{g\in \mathcal{G}} \min_{p\in \mathcal{P}} d(p,g)\right)$\\ 
				Absolute Volume Difference & $AVD=\frac{|\mathcal{V}_p - \mathcal{V}_g|}{\mathcal{V}_g}\times 100$\\
				\hline
				\label{tab_1}    
		\end{tabular}}
	\end{table}
	
	\subsubsection{Performance metrics}
	The performance metrics are the key factors to evaluate and compare the segmentation performance of the models. Due to the unavailability of the standard metrics, each system requires an appropriate and different selection of metrics that can quantify time, computational and memory space requirements and overall performance~\citep{fenster2006evaluation}. Table~\ref{tab4} presents the most popular evaluation metrics that are utilized to analyse the performance in BIS models. In BIS, mostly the datasets are imbalanced i.e. the number of pixels/voxels concerning the target region (region of interest) are relatively less than the dark pixels/voxels (background region), due to which the metrics such as accuracy, which are best suited for a balanced distribution of data samples, are not recommended for BIS evaluation of the models. Among the discussed metrics intersection-over-union (IoU or Jaccard index) and dice similarity coefficient are the most widely used evaluation metrics in BIS for various modalities. More details can be found in the recent review articles~\citep{haque2020deep, minaee2020image}.
	
	\begin{table}[!b]
		\centering
		\caption{Summary of the most widely used loss functions for biomedical image segmentation with respect to the predicted mask ($\mathcal{P}$) and ground truth mask ($\mathcal{G}$), $\alpha$ and $\gamma$ as constants, $h$ is Hausdorff distance and $d$ is the operator for Euclidean distance.}
		\label{tab5}
		
		\begin{tabular}{llP{1.1in}}
			\hline
			\textbf{Type} & \textbf{Objective functions} & \textbf{Usecase}\\
			\hline
			\multicolumn{1}{l}{\multirow{7}{*}{Distribution}} & $\mathcal{L}_{BCE}=-(glog(p)+(1-g)log(1-p))$ & Balanced distribution of data\\
			& $\mathcal{L}_{WCE}=-(\alpha .glog(p)+(1-g)log(1-p))$ & Skewed dataset\\
			& $\mathcal{L}_{BaCE}=-(\alpha glog(p)+(1-\alpha)(1-g)log(1-p))$  & Skewed dataset\\
			& $\mathcal{L}_{Focal}=\left \{
			\begin{aligned}
				& -\alpha(1-p)^\gamma log(p), && \text{if}\ g=1 \\
				& -(1-\alpha)(p)^\gamma log(1-p), && \text{otherwise}
			\end{aligned} \right.$ & Focuses on hard samples\\
			\hline
			\multicolumn{1}{l}{\multirow{8}{*}{Region}} & $\mathcal{L}_{DSC}=1-\frac{2gp+1}{g+p+1}$ & Widely used for segmentation\\
			& $\mathcal{L}_{IoU}=1-\frac{gp}{g+p-gp}$ & Widely used for segmentation\\
			& $\mathcal{L}_{SS}=\alpha*\mathrm{sensitivity}+(1-\alpha)*\mathrm{specificity}$ & Focuses to improve true positive rate\\
			& $\mathcal{L}_{Tversky}=1- \frac{1+gp}{1+gp+\alpha(1-g)p+(1-\alpha)g(1-p)}$ & Introduces weights for false predictions\\
			\hline
			\multicolumn{1}{l}{\multirow{5}{*}{Boundary}} & $\mathcal{L}_{HD}=\frac{1}{N} \sum_{i=0}^{N} \Big((p_i-g_i)^2.( h_{pi}^{2} + h_{gi}^{2} ) \Big)$ & Widely used for segmentation\\
			& $\mathcal{L}_{SA}=-\sum_i CE(p_i,g_i)-\sum_i{\alpha_i}d(\mathcal{P},\mathcal{G}) CE(p_i,g_i)$ & Focuses to segment boundaries of the regions\\
			\hline
			\multicolumn{1}{l}{\multirow{5}{*}{Compound}} & $\mathcal{L}_{Combo}=\alpha \mathcal{L}_{BaCE}(g,p)-(1-\alpha)\mathcal{L}_{DSC}(g,p)$ & Leverages features of $BaCE$ and $DSC$ for skewed data\\
			& $\mathcal{L}_{EL}=\alpha_{DSC}e^{(-ln(\mathcal{L}_{DSC})^\gamma)}+\alpha_{CE}\,e^{(-ln(\mathcal{L}_{CE})^\gamma)}$ & Focuses on less accurate predictions\\
			\hline
			\multicolumn{3}{p{4.5in}}{$BCE$ - binary cross-entropy, $WCE$ - weighted cross-entropy, $BaCE$ - balanced cross-entropy, $DSC$ - dice similarity coefficient, $IoU$ - intersection-over-union, $SS$ - sensitivity-specificity, $HD$ - Hausdorff distance, $SA$ - shape-aware, $EL$ - exponential-logarithmic.}
		\end{tabular}
		
	\end{table}
	
	\subsubsection{Loss functions}
	The loss functions or objective functions drive the training procedure of the deep learning models. For the BIS task, loss functions are tuned to alleviate the above discussed class imbalance problem by refining the distributions of the training data. With each dataset introducing its complexities and challenges, the loss functions are grouped into four categories based on the distribution, region, boundary and hybrid~\citep{ma2020segmentation}, as shown in Table~\ref{tab5}, along with their respective usecases. For ease in representation, the loss functions are summarized for the semantic segmentation scenario, where the number of classes is limited to two (background and target region). The effect of these loss functions for biomedical image segmentation using various modalities over nnU-Net model \citep{isensee2021nnu} is explored by \cite{nasalwai2021addressing}, and also proposed an accelerated tversky loss (ATL) function to achieve faster model training or convergence. 
	
	\section{U-Net variants for medical imaging}
	\label{sec3}
	\begin{figure}[!b]
		\centering
		\includegraphics[scale=0.45] {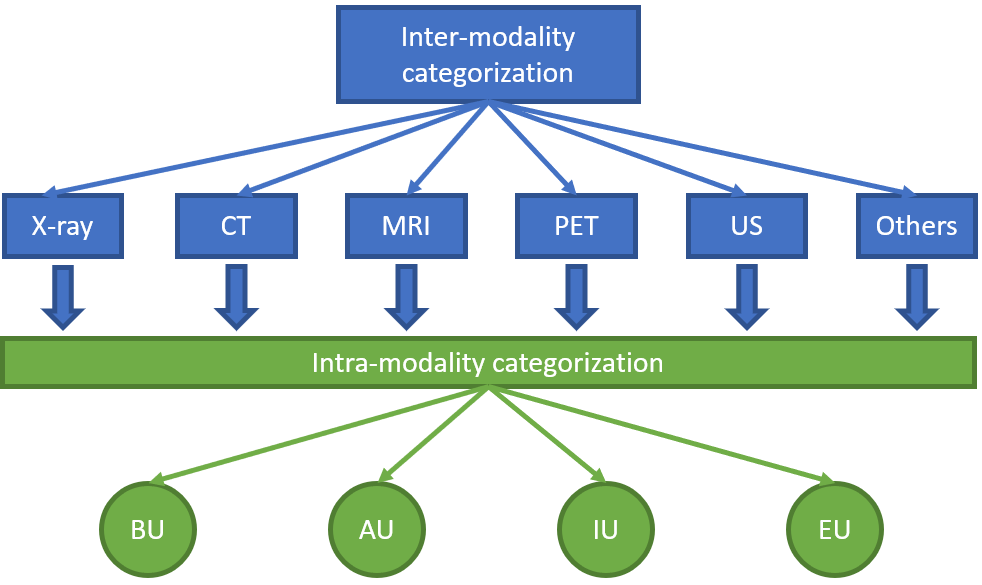}
		\caption{Categorization scheme for U-Net variants.}
		\label{fig7}
	\end{figure}
	The numerous development in medical imaging acquisition systems and deep learning technologies have resulted in the rise of usage frequency of modalities for computer-aided diagnosis. Despite vanilla U-Net being super-efficient in the ISBI cell tracking challenge, there is still a void to fill with improvements in certain aspects. The most apparent problem in the vanilla U-Net is that the learning may slow down in deeper layers of the U-Net model which increases the possibility of the network ignoring the layers representing abstract features of the target structure. This slack in the learning process is due to the generation of diluted gradients in the deeper layers. Another major issue is concerned with the localized convolutions which tend to limit the capability of the model to efficiently capture global and long-range dependencies. Furthermore, the distinct challenges (discussed in a later section) introduced in performing segmentation using different modalities needs to be addressed; however, it is not optimal with vanilla U-Net. Following this context, various U-Net variants are proposed to improve the segmentation performance across vivid modalities. To establish a better understanding of these variants, the present review performs: 1) inter-modality categorization - to show variation in the segmentation approaches across the different modalities (X-ray, CT, MRI, PET and ultrasound), and 2) intra-modality categorization -  to group each U-Net variant within the same modality based on its most profound technical contribution (better U-Nets, attention U-Nets, inception U-Nets and ensemble U-Nets), as shown in Fig.~\ref{fig7}. Within each modality, a similar type of categorization is performed to better distinguish the type of approaches introduced for each modality. In the case where a U-Net variant uses multiple modifications, then based on its most profound enhancement or technical contribution it is added in that category, e.g. RCA-IUnet model \citep{punn2021rca} for breast cancer segmentation using ultrasound imaging, uses residual cross-spatial attention and inception convolutions, is categorized as an attention U-Net variant under ultrasound modality. Each of the intra-modality categories are described as follows:  
	
	\begin{itemize}
		\item 
		Better U-Nets (BU): This category consists of U-Net models that are better than raw U-Net model by slight modifications such as integrating with FCN or SegNet models, transfer learning, dense or residual blocks, multi-stage training, multi-tasking, etc.
		\item
		Attention U-Nets (AU): These are the approaches with variation in the attention mechanism of the feature maps to filter the relevant features such as spatial and channel attention, mixed attention, non-local attention, etc.
		\item
		Inception U-Nets (IU): These are the approaches that use multi-scale feature fusion strategies to effectively learn the feature representations.
		\item
		Ensemble U-Nets (EU): These are the approaches that use multiple models or sub-models with or without the other enhancements to improve the segmentation performance.
	\end{itemize}

	Hence, for faster and efficient computer-aided diagnosis practices, the following sections present wide varieties of U-Net based approaches for biomedical image segmentation using various modalities. Table~\ref{tab6} summarizes the various U-Net variants reviewed in the following sections. 
	
	\begin{table}
		\caption{Summary of popular U-Net variants for BIS.} \label{tab6}
		\resizebox{\columnwidth}{!}{\begin{tabular}{P{1in}P{0.8in}P{0.5in}P{0.15in}P{0.10in}P{0.12in}P{0.12in}P{0.4in}P{1.4in}}
				\hline
				\textbf{Author} & \textbf{U-Net variant} & \textbf{Modality} & \textbf{TL} & \textbf{SL} & \textbf{Pr} & \textbf{Po} & 
				\textbf{Category} &
				\textbf{Description} \\
				\hline
				\cite{dong2017automatic}  & Modified U-Net & MRI & - & \checkmark & \checkmark & - &  BU & FCN based U-Net \\
				\cite{rashid2018fully}  & Modified U-Net & X-ray & - & \checkmark & \checkmark & \checkmark & BU & FCN based U-Net \\
				\cite{frid2018improving}  & Modified U-Net & X-ray & \checkmark & - & \checkmark & - & BU & U-Net with pre-trained VGG-16 encoder \\
				\cite{que2018cardioxnet}  & CardioXNet framework & X-ray & - & \checkmark & \checkmark & \checkmark & EU & Two parallel U-Net models with binary contours\\
				\cite{oktay2018attention}  & Attention U-Net & CT & - & \checkmark & \checkmark & - & AU & Attention skip-connections\\
				\cite{kohl2018probabilistic}  & Probabilistic U-Net & CT & - & \checkmark & - & - & EU & U-Net with conditional variational autoencoder \\
				\cite{tong2018improved}  & Improved U-Net & CT & - & \checkmark & \checkmark & - & BU & Mini-residual connections within encoder-decoder phases \\
				\cite{janssens2018fully}  & Two stage U-Net model & CT & - & \checkmark & \checkmark & - & EU & 3D FCN LocalizationNet followed by SegmentationNet \\
				\cite{kumar2018u}  & U-SegNet & MRI & \checkmark & - & \checkmark & - & BU & Integration of skip connections with SegNet \\
				\cite{kermi2018deep}  & Residual U-Net & MRI & - & \checkmark & \checkmark & - & BU & Residual blocks between two convolution layers \\
				\cite{chen2018s3d}  & S3DU-Net & MRI & - & \checkmark & \checkmark & - & IU & U-Net with spatiotemporal separable convolution \\
				\cite{blanc2018automatic}  & Vanilla 3D U-Net & PET & - & \checkmark & \checkmark & \checkmark & BU & CNN based 3D U-Net \\
				\cite{zhao2018tumor}  & 3D FCN & PET & - & \checkmark & \checkmark & - & IU & 3D FCN multi-modal fusion network \\
				\cite{almajalid2018development}  & U-Net + SRAD & US & - & \checkmark & \checkmark & \checkmark & BU & Base U-Net with speckle reducing anisotropic diffusion \\
				\cite{wang2018simultaneous}  & cU-Net & US & - & \checkmark & \checkmark & - & EU & Classification and segmentation U-Net \\
				\cite{alom2018recurrent}  & R2U-Net & Multi-modality & - & \checkmark & \checkmark & \checkmark & BU & Recurrent Residual convolutional neural network based on U-Net (R2U-Net)\\ 
				\cite{zhou2018unet}  & UNet++ & Multi-modality & - & \checkmark & \checkmark & - & BU & Nested U-Net model\\
				\cite{subramanian2019automated}  & CVC framework & X-ray & \checkmark & - & - & - & EU & Two parallel U-Net models with spatial priors and pre-trained NN-RF\\
				\cite{li2019attention}  & U-Net based framework & X-ray & \checkmark & - & \checkmark & \checkmark & AU & SE and residual based attention CNN \\
				\cite{dong2019automatic}  & U-Net-GAN & CT & - & \checkmark & \checkmark & - & EU & U-Net act as a generator and FCN as discriminator network \\
				\cite{liu2019liver}  & GIU-Net & CT & - & \checkmark & \checkmark & \checkmark & EU & Deeper U-Net model with graph cut algorithm\\
				\cite{man2019deep}  & GAU-Net & CT & \checkmark & - & \checkmark & - & AU & Deformable geometry-aware U-Net with deep Q learning \\
				\cite{seo2019modified}  & mU-Net & CT & - & \checkmark & - & - & AU & Object dependent filters in skip connections \\
				\cite{hiasa2019automated}  & Bayesian U-Net & CT & - & \checkmark & \checkmark & \checkmark & EU & Cascaded U-Net and Bayesian U-Net models \\
				\cite{song2019u}  & U-NeXt & CT & - & \checkmark & \checkmark & - & AU & U-Net model with attention blocks, SkipSPP and dense convolutions \\
				\cite{rundo2019use}  & USE-Net & MRI & - & \checkmark & \checkmark & \checkmark & AU & U-Net model with the squeeze-and-excitation blocks \\
				\cite{wang2019msu}  & MSU-Net & MRI & - & \checkmark & \checkmark & - & IU & Multiscale statistical U-Net \\
				\cite{dong2019synthetic}  & DAU-Net & MRI & - & \checkmark & - & - & AU & Deep attention U-Net with deep supervision \\
				\cite{wang2019deeply}  & 3D DSD-FCN & MRI & - & \checkmark & \checkmark & \checkmark & EU & 3D FCN with deep supervision and group dilation \\
				\cite{guo2019gross}  & 3D Dense U-Net & PET & - & \checkmark & \checkmark & - & IU & 3D U-Net with dense convolution fusion blocks \\
				\cite{yang2019robust}  & DPU-Net & US & - & \checkmark & \checkmark & - & IU & Dual path U-Net with parallel multi-branch encoding and decoding\\
				\cite{li2019automatic}  & DU-Net & US & - & \checkmark & \checkmark & \checkmark & BU & Dense convolution U-Net \\
				\hline
				\multicolumn{2}{l}{\textit{continue to the next page}}
		\end{tabular}}
	\end{table}	
	
	\begin{table*}[!h]
		\caption*{}
		\resizebox{\columnwidth}{!}{\begin{tabular}{P{1in}P{0.8in}P{0.5in}P{0.15in}P{0.10in}P{0.12in}P{0.12in}P{0.4in}P{1.4in}}
				\hline
				\cite{lin2019semantic}  & SSU-Net & US & - & \checkmark & \checkmark & - & AU & Semantic-embedding and shape-aware U-net \\
				\cite{azad2019bi}  & BCDU-Net & Multi-modality & - & \checkmark & \checkmark & - & BU & Bi-directional ConvLSTM U-Net with densley connected convolutions\\
				\cite{gu2019net}  & CE-Net & Multi-modality & \checkmark & - & \checkmark & - & IU & U-Net based context encoder network\\
				\cite{abedalla20202st}  & 2STU-Net & X-ray & \checkmark & - & \checkmark & \checkmark & BU & Two stage U-Net with pre-trained ResNet-34 model \\
				\cite{zhang2020dual}  & DEFU-Net & X-ray & - & \checkmark & \checkmark & - & IU & U-Net with encoder fusion of dense and inception CNN \\
				\cite{wang2020mdu}  & MDU-Net & X-ray & \checkmark & - & \checkmark & - & BU & Multi-task dense connection U-Net \\
				\cite{park2020fully}  & 3D U-Net & CT & - & \checkmark & \checkmark & \checkmark & BU & 3D U-Net with segmentation error correction\\
				\cite{fan2020ma}  & MA-Net & CT & - & \checkmark & \checkmark & - & AU & U-Net based multi-scale attention model \\		
				\cite{dong2020deu}  & DeU-Net & MRI & - & \checkmark & \checkmark & - & AU & 3D deformable attention U-Net\\
				\cite{punn2020multi}  & 3D inception U-Net & MRI & - & \checkmark & \checkmark & - & EU & 3D inception U-Net with modality fusion \\
				\cite{lu2020automatic}  & Modified U-Net & PET & \checkmark & - & - & \checkmark & BU & U-Net with pre-trained VGG-19 encoder\\
				\cite{leung2020physics}  & Modified U-Net & PET & \checkmark & - & \checkmark & - & EU & Physics guided minimal U-Net with dropout regularization \\
				\cite{dunnhofer2020siam}  & Siam-U-Net & US & - & \checkmark & \checkmark & - & EU & U-Net with siamese tracking framework\\
				\cite{zhang2020multi}  & AU-Net & US & - & \checkmark & \checkmark & - & AU & Attention guided U-Net with total variation regularization \\
				\cite{BYRA2020102027}  & SKU-Net & US & - & \checkmark & \checkmark & - & AU & Attention based selective kernel U-Net\\
				\cite{punn2020inception}  & IU-Net & Histopathol-ogical & - & \checkmark & \checkmark & - & IU & Inception U-Net model with hybrid spectral pooling \\
				\cite{ibtehaz2020multiresunet}  & MR-UNet & Multi-modality & - & \checkmark & \checkmark & - & IU & MultiResUNet with multiple inception based skip connections \\
				\cite{wang2020non} & NL-Unet & Multi-modality & - & \checkmark & - & - & AU & Non-local Unet with global context aggregation\\
				\cite{xia2021mc}  & MC-Net & CT & - & \checkmark & \checkmark & - & IU & Multi-scale context extraction with residual attention \\
				\cite{li2021automatic}  & MSA-Unet & MRI & - & \checkmark & \checkmark & - & AU & U-Net with dual branch multi-scale attention \\
				\cite{fu2021multimodal}  & MSAM-Net & PET & - & \checkmark & \checkmark & - & AU & Multi-modal spatial attention network \\
				\cite{punn2021rca} & RCA-IUnet & US & - & \checkmark & - & - & AU & Residual cross-spatial attention guided inception U-Net model \\
				\cite{cao2021swin} & Swin-Unet & Multi-modality & \checkmark & - & - & - & EU & Unet-like pure transformer network \\
				\cite{wang2021mixed} & MTM-Unet & Multi-modality & - & \checkmark & - & - & EU & U-Net with mixed transformer module\\
				\cite{isensee2021nnu}  & nnU-Net & Multi-modality & - & \checkmark & \checkmark & \checkmark & EU & Self-adapting no-newU-Net Framework\\
				\hline
				\multicolumn{9}{p{6in}}{TL - Transfer learning, SL - Scratch learning, Pr - Pre-processing, Po - Post-processing}\\
		\end{tabular}}
	\end{table*}
	
	\subsection{X-ray}
	In radiology, X-ray imaging is utilized as a diagnostic procedure of human bones and tissues. X-ray possesses the properties of penetrability, photographic effect and fluorescence effect. Human body tissues vary in density and thickness due to which X-rays are absorbed with different degrees, resulting in black and white contrast images~\citep{bercovich2018medical}. The wide and easy availability of X-ray imaging has encouraged the research community to contribute towards smart diagnosis systems.
	
	\subsubsection{Better U-Nets}
	
	The segmentation of lungs from chest X-ray (CXR) imaging is a crucial step for any CAD system. Following this, \cite{rashid2018fully} exploits the potential of U-Net model to generate the segmentation masks of the lungs from CXR images, where the produced masks are iteratively refined with post-processing techniques such as flood fill algorithm and morphological operations. Significant improvement is observed as compared to traditional segmentation approaches such as adaptive region growing, edge detection, statistical shape models, etc., over multiple datasets. To further improve segmentation performance, \cite{frid2018improving} employed a pre-trained VGG-16 model in the encoder phase, where the decoder or the expansion phase uses upsampling and standard convolution operations sequentially for multi-class segmentation involving anatomical structures like lungs field, heart and clavicles in chest X-ray samples. While training, the pre-trained weights are fine-tuned to better extract or encode the desired features of the target classes. Unlike the approach by \cite{rashid2018fully}, this model with transfer learning achieved promising results without any post-processing overhead. Besides, the authors also analysed the proposed model with multiple loss functions like $DSC$, $IoU$, Tversky and $BCE$, where the use of $DSC$ produced the best results.
	
	In another work, \cite{abedalla20202st} proposed a deep learning framework 2STU-Net to perform segmentation of pneumothorax (collapsed lung) in the CXR samples. It comprises a state-of-the-art residual network (ResNet-34) that is pre-trained on the ImageNet dataset and arranged in the U-Net topology. Similar to the work by \cite{frid2018improving}, the encoder is built with ResNet-34~\citep{he2016deep} by removing the last layers, whereas the decoder follows standard blocks of CNN with upsampling. Initially, the data is pre-processed to produce images of dimensions 256$\times$256 and 512$\times$512 for 2 stage training scheme and multi-scale feature learning. The ResNet34U-Net is first trained with lower resolution images and later the same model is fine-tuned (keeping previously learned weights as initial weights) to adapt high resolution images. The authors also utilized stochastic weight averaging (SWA) and test-time augmentation (TTA) techniques to improve the test results. The significance of 2 stage training is justified with the faster convergence of the second training stage and better segmentation results, thereby highlighting the effectiveness of multi-scale feature representation learning. However, the overall training overhead is increased due to two stage training. \cite{wang2020mdu} synthesized a CXR dataset annotated with clavicles, anterior ribs, posterior ribs and bones, on which a multi-task dense connection U-Net (MDU-Net) is trained for multi-class segmentation. A feature separation network is introduced for multi-label segmentation where a pixel value is associated with more than one label e.g. the pixels in the overlapped regions of anterior and posterior ribs have multiple tags. For every CXR image, multiple masks are generated concerning different annotations, thereby multiple networks are trained to generate the corresponding mask. The implication of increased training time is addressed with the help of transfer learning, where the network uses a pre-trained DenseNet201~\citep{huang2017densely} model for feature extraction. However, due to the 2D projection of X-ray imaging,  each annotated mask also covers features representing other masks categories which may deviate the network from learning the class specific feature representations.
	
	\subsubsection{Attention U-Nets}
	Motivated by the success of squeeze-and-excitation network (SENet)~\citep{hu2018squeeze} to suppress the irrelevant features, \cite{li2019attention} proposed an attention guided deep learning framework divided into three components: preprocessing, region of interest (RoI) segmentation with transfer learning followed by pneumonia detection model. In the preprocessing stage, apart from the trivial processes like resizing, the authors synthesized the adversarial samples to gain attention of the model towards pneumonia. The pneumonia infected area is erased by replacing it with an average pixel value of the image and then labelled as non-pneumonia, which helped to distinguish between noise and relevant data. To further suppress the background interference, authors adopted the approach proposed by \cite{rashid2018fully} to perform the lungs segmentation followed by post-processing with conditional random fields. The segmented, original and synthesized images together form the training and validation set for the pneumonia segmentation network. The network follows SENet design in which SE-ResNet34 is utilized as a backbone architecture. The proposed framework tends to learn the pneumonia features effectively and achieves a significant reduction in the false positive predictions with an FPR value of 0.19, in contrast to mask R-CNN~\citep{he2017mask} and RetinaNet~\citep{lin2017focal} on RSNA challenge; however, the overall framework relies heavily on the pre-processing and post-processing of data, thereby limiting its usability across multiple datasets.
	
	\subsubsection{Inception U-Nets}
	In another U-Net variant, \cite{zhang2020dual} proposed a DEFU-Net model that uses the fusion of dual encoder models to better extract the spatial features and a standard decoder network with upsampling. The dual encoder network is equipped with a densely connected recurrent convolutional (DCRC) neural network (inspired from DenseNet~\citep{huang2017densely} and R2U-Net~\citep{alom2018recurrent}) and dilated inception convolution neural network (inspired from GoogleNet~\citep{szegedy2015going}), where the output from each layer is merged by addition operation which is later concatenated with the corresponding decoder layer. The DCRC aids in extracting high level features, whereas the inception block facilitates to increase the network width and improve the horizontal feature representation using various receptive fields with dilated convolutions. The advantage of using dilated convolutions is that it tends to increase the receptive field without changing the number of training parameters~\citep{yu2016multiscale}. The significance of each module of the network is established by achieving considerable improvements over several U-Net variants such as residual U-Net~\citep{he2016identity}, BCDU-Net~\citep{azad2019bi}, R2U-Net and attention R2U-Net~\citep{alom2018recurrent}, etc. with dice score of 0.97 on the chest X-ray dataset.
	
	\subsubsection{Ensemble U-Nets}
	With cardiomegaly being one of the most common inherited cardiovascular diseases, \cite{que2018cardioxnet} proposed a CardioXNet framework to identify and localize the cardiomegaly present in the chest X-ray images. CardioXNet is equipped with two parallel U-Net models to generate the segmentation masks for cardiac and thorax respectively, that follows typical CNN architecture in contraction and expansion paths. To address the limited availability of the data samples, authors utilized data augmentation strategies such as rotation, zooming, shearing, etc. Due to the possibility of the presence of noise in the output masks, post-processing is applied to keep the binary contours that represent the largest area. Later, the processed output mask is utilized to compute the cardiothoracic ratio defined as $CTR=(L+R) / (T)$, where $L$ and $R$ indicates the maximum distances from the center to the left and right farthest boundaries of the heart region, and $T$ is the maximum horizontal distance between the lungs boundaries. The $CTR$ value is then utilized to determine the cardiomegaly from the generated masks. In another approach, \cite{subramanian2019automated} proposed an automated system involving two U-Net models, where the output features are exploited to identify the type of central venous catheters (CVC) as peripherally inserted central catheters (PICC), internal jugular (IJ), subclavian and Swan-Ganz catheters. The first U-Net model is utilized for CVC segmentation by using the exponential logarithmic loss to address the class imbalance problem, whereas the other U-Net model tends to extract the anatomical structures to distinguish the ambiguous classes such as PICC and subclavian lines. Clinicians manually annotated the CVCs to obtain the signature spatial priors which undergo pixel-wise multiplication with the segmentation output. Later, the produced output is fed to the pre-trained neural network random forest (NN-RF) classifier to distinguish the type of CVC. This hybrid combination of segmentation and classification achieved promising results on the NIH database.	
	
	\subsection{Computed tomography}
	Computed tomography imaging is based on the principle of utilizing the series of the system of rotating X-rays to develop cross-sectional images or series of slices of bones, blood vessels and soft tissues of the body~\citep{bercovich2018medical}. In contrast to plain X-ray imaging, CT scans provide rich information with high quality images. This is generally utilized to examine people with serious injuries or diseases like trauma, tumors, pneumonia, etc., and also to plan medical, surgical or radiation treatment. Hence, various deep learning based approaches are developed for faster diagnosis and treatment using CT imaging.
	
	\subsubsection{Better U-Nets}
	\cite{tong2018improved} proposed a U-Net framework for lung nodule segmentation, where mini residual connections are introduced within the encoder and decoder phases to address the vanishing gradient problem. The algorithm initiates with the process of generating the segmentation of lung parenchyma with morphological operations and removal of irrelevant features. The segmented lung parenchyma images are divided into 64$\times$64 slices along with the input images. Finally, the improved U-Net model is trained and validated against the preprocessed dataset for segmenting the pulmonary nodules. The authors evaluated the approach on LUNA2016 dataset against various models and achieved promising results with a dice score of 0.74; however, the samples of pulmonary nodules were very limited and the approach also lacked the 3D volumetric analysis. Recently, \cite{park2020fully} utilized a 3D U-Net model to segment the lung lobe regions while also addressing the miss-detection of the lobar fissure. Initially, the volumetric CT scans are preprocessed with thresholding to identify lungs parenchyma, and region growing techniques~\citep{leader2003automated} to separate overlapping left and right lung regions. Later, these lobe segmentations are generated with the help of the 3D U-Net model, where the segmentation results are further refined with the upsampling and segmentation error correction. The authors utilized CT volumes from multiple centres (hospitals) to evaluate the model performance while achieving significant improvements.
	
	\subsubsection{Attention U-Nets}
	When the target is the segmentation of the internal organs, then models adopting the attention mechanism help to focus the network on regions of interest. \cite{oktay2018attention} proposed a novel attention gate based U-Net framework to focus on pancreas regions and generate the corresponding segmentation masks. The attention approach tends to suppress irrelevant features and highlight the prominent features corresponding to the target regions. The authors utilized the FCN with U-Net connectivity, where the skip connections are loaded with these attention filters. Inspired from the work of \cite{shen2017disan}, each pixel is associated with a gating vector to determine the regions to focus. The incorporation of this attention mechanism allowed the authors to achieve significant improvements in the segmentation results over other approaches. The performance of this model could easily be improved by incorporating transfer learning, multi-stage training, etc. To exploit the potential of attention mechanism, \cite{seo2019modified} proposed a modified U-Net (mU-Net) framework that addressed the classical problems associated with the standard U-Net model concerning skip connection~\citep{han2018framing} and pooling operation (loss of spatial information). In the mU-Net, the standard skip connections are replaced by object-dependent filters to dynamically filter the feature maps based on the object size, where features concerning the small objects are preserved by blocking the deconvolution path and in the case of large objects, feature maps indicating boundary information is propagated to avoid duplication. The authors verified the effectiveness of adaptive filters to preserve the features using the permeation rate while achieving the $DSC$ values of 0.98 and 0.89 on the liver and liver-tumor segmentation respectively. The approach could be extended for 3D volumetric analysis, where computation cost can be addressed by modifying operation schemes such as depthwise separable convolution instead of standard convolution. These object-dependent filters could easily be integrated with other networks and modalities for segmentation.
	
	\cite{song2019u} proposed a U-NeXt model to segment CT images of gallstones, which is one of the common and frequently occurring diseases worldwide. The U-NeXt model is equipped with the attention up-sampling blocks, spatial pyramid pooling~\citep{he2015spatial} of skip connections (SkipSPP), and multi-scale feature extraction with the series of convolution layers along with the dense connections. The overall architecture design is similar to U-Net++ model~\citep{zhou2018unet++} with slight variation in connections, convolution and pooling operations. The authors trained and evaluated the model on the proposed dataset with 5,350 images using deep supervision and reported improvement in IoU by 7\% over baseline biomedical image segmentation models. However, for complex target structures, the network produces soft edges in the mask. To address this issue, a deep Q network (DQN)~\citep{mnih2015human} driven approach is proposed by \cite{man2019deep} that uses deformable U-Net to efficiently generate the segmentation mask of the pancreas from CT scans with the extraction of its contextual information and anisotropic features. Initially, the 3D volumes are split into axial, coronal and sagittal 2D slices for each of which, DQN-based deep reinforcement learning (DRL) agents tend to localize the pancreas to form RoI slices. These slices are fed to the deformable U-Net models and finally, based on the majority voting scheme 3D segmentation mask is generated. The deformable U-Net~\citep{dai2017deformable} follows standard encoder-decoder architecture, where convolution operations are replaced with deformable convolutions (DC). In DC, the regular convolution operation is accompanied by another convolution layer to learn 2D offset for each pixel. It leverages the deep network's ability to learn the required receptive field rather than being fixed for segmenting the regions having varying geometrical structures. This can also be understood as a learnable dilated convolution.
	
	Unlike other U-Net variants that applies multi-scale feature fusion, \cite{fan2020ma} recently proposed a multi-scale attention U-Net model that uses a self attention scheme for adaptive feature extraction. The self attention design comprises position-wise attention block (PAB - installed on bottleneck layer) and multi-scale fusion attention block (MFAB - installed on every stage of encoder path), where PAB captures feature interdependencies in spatial dimension and MFAB captures the channel dependencies for any feature map. The MA-Net is trained and evaluated on the 2017 LiTS challenge and achieved a $DSC$ score of 0.96 and 0.75 for liver and liver-tumor segmentation respectively. However, the results are not as promising as achieved using mU-Net model~\citep{seo2019modified}.  
	
	\subsubsection{Inception U-Nets}
	Recently, \cite{xia2021mc} proposed MC-Net that uses a multi-scale context extraction module with a context residual attention approach to model the local and global semantic information of the target regions using CT imaging and alleviate the problem of not capturing the long-range dependencies by most of the U-Net models. Overall, MC-Net is built with a multi-feature extraction module (MIE) to obtain multi-scale feature maps similar to inception network \citep{szegedy2017inception}, context information extraction (CIE) with parallel dilated convolutions and residual attention enabled skip connections. Though the model achieved promising results over multiple CT datasets; however, the diversity of multi-scale feature extraction is limited to a feature map at a single level, which could be improved by considering feature maps across different encoding layers.
	
	\subsubsection{Ensemble U-Nets}
	\cite{janssens2018fully} proposed a cascaded 3D FCN based deep learning model consisting of \enquote{LocalizationNet} and \enquote{SegmentationNet} to estimate the bounding box (RoI) and generate volumetric segmentation masks of lumbar vertebrae respectively. The LocalizationNet comprises a 3D FCN regression model which is trained to regress the displacement vectors associated with a voxel, representing diagonal corners of the rectangular box. The localized information is fed to SegmentationNet comprising an FCN 3D U-Net model to produce a segmentation mask for lumbar vertebrae. This two stage approach exhibited significant improvement over the existing approaches but with the overhead computations of two dedicated models. In the real world scenario, modalities may suffer from inherent ambiguities that coagulate the actual nature of the disease. Following this, \cite{kohl2018probabilistic} introduced a probabilistic U-Net framework that combines the standard U-Net model with conditional variational autoencoder (CVAE). For a sample image, CVAE generates diverse plausible hypotheses from a low-dimensional latent space which are fed to U-Net to generate the corresponding segmentation mask. It is shown that the model can generate diverse segmentation samples, given the ground-truth delineation from multiple experts. The trained model is evaluated on LIDC-IDRI and Cityscapes datasets which outperformed other approaches in reproducing the segmentation probabilities and masks. Inspired by this work many other variants have been developed to capture the uncertainties, e.g.~\citep{raghu2019direct, baumgartner2019phiseg, tanno2019learning}.
	
	Motivated by the success of adversarial techniques, \cite{dong2019automatic} proposed a U-Net-GAN framework in which a set of U-Nets is trained as a generator to produce organs-at-risk (OARs) segmentation and FCN as a discriminator to distinguish segmented masks from the ground-truth masks. The generator and discriminator networks followed adversarial training, where each network competes to achieve optimal segmentation masks of OARs. The model achieved satisfactory improvements at the cost of heavy computations and resource requirements, moreover, the model struggles in the presence of complex structures. In another work, \cite{liu2019liver} proposed a liver CT image segmentation framework named GIU-Net, inspired by the supervised interactive segmentation approach named, graph cut~\citep{boykov2006graph}. The improved U-Net model is designed with increased depth to better extract semantic features that are trained to generate the segmentation mask of the liver regions. Later, to further refine the segmentation results, a slice covering the maximum liver region is used as an initial slice to generate graph cut energy function followed by the maximum flow minimum cut algorithm. The process is then repeated for all the slices to generate a complete sequence of precise and stable segmentation masks with smother boundaries.
	
	In another application, a Bayesian CNN with U-Net model and Monte Carlo (MC) dropout is introduced by \cite{hiasa2019automated} for automated muscle segmentation from CT imaging for musculoskeletal modelling. The design comprises two cascaded U-Net models, where first is standard U-Net that localizes the skin surface and later individual muscles (21 muscles) are segmented with Bayesian U-Net~\citep{kendall2015bayesian} that uses MC dropout based on the structure-wise uncertainty, predictive structure-wise variance (PSV) and predictive dice coefficient (PDC). Besides, the authors employed an active learning method to produce segmentation and uncertainty from the unlabeled data, where the high uncertain data are relabeled manually by experts while other data is directly used as training data. The authors achieved promising results; however, the data samples were very limited which limits the diversity of the model.

	\subsection{Magnetic resonance imaging}
	Magnetic resonance imaging is synthesized by using the principles of nuclear magnetic resonance (NMR)~\citep{morris2018magnetic}. It is utilized in radiology to visualize the anatomy and physiological process of the body organs. It uses a large magnetic field and radio waves to create detailed images of organs and tissues within the body. Based on the different attenuation values of the tissues e.g. T1-weighted (T1), fluid attenuation inversion recovery (FLAIR), Dixon, etc.,  the electromagnetic waves emitted from the gradient magnetic field is detected using the applied strong magnetic field by which the position and type of the nucleus can be drawn inside the object. Unlike the X-rays, CT scans and PET scans; MRI scans do not involve the usage of ionizing radiations. 
	
	\subsubsection{Better U-Nets}
	MRI is mostly utilized in computer-aided diagnosis systems involving brain tumor segmentation. Inspired from the BraTS 2015 challenge, \cite{dong2017automatic} analysed the potential of FCN based U-Net model for brain tumor segmentation via MRI sequences, where the authors achieved significant improvement over the traditional segmentation approaches. SegNet is another model that is most widely used for semantic segmentation~\citep{badrinarayanan2017segnet}. Following this, \cite{kumar2018u} proposed a hybrid approach, U-SegNet, by integrating skip connections into the base SegNet model. This enabled the model to efficiently identify the tissue boundaries concerning the white matter (WM), gray matter (GM), and cerebro-spinal fluid (CSF). The authors achieved significant improvement over the base SegNet and U-Net models with $DSC$ value of 0.90 on IBSR-18 dataset.
	
	In recent years, to improve the biomedical image segmentation results, multi-modality fusion (MMF)~\citep{james2014medical} approaches are utilized. The fused scans are rich in information and offer multi-dimensional features. In this context, \cite{kermi2018deep} proposed a modified U-Net model to segment the whole tumor and intra tumor regions like enhancing tumor, edema and necrosis affected with high grade glioma (HGG) and lower grade glioma (LGG) following from the BraTS 2018 challenge. The authors fused the T1, T2, T1c and FLAIR modalities and resized them to form the input feature map with rich tumor information. In the modified model, residual blocks~\citep{he2016identity} are added between two convolution blocks and the max-pooling operation is replaced with the strided convolutions~\citep{ayachi2018strided}. The model is trained and evaluated with fused modalities to obtain the multi-class segmentation masks. Though the authors achieved good results but lacked the 3D volumetric analysis. In another application, skull stripping is an essential step to study brain imaging, where \cite{hwang20193d} proposed to utilize a standard 3D U-Net model to automate the process of skull stripping (brain extraction) from T1 MRI scans for faster diagnosis and treatment. The training is carried with dice loss and adam optimizer on neurofeedback skull-stripped (NFBS) dataset. The authors achieved a dice value of 0.99; however, the comparative study is limited to brain surface extractor (BSE) and robust brain extraction (ROBEX) algorithms.
	
	\subsubsection{Attention U-Nets}
	With the introduction of modality transformations, \cite{dong2019synthetic} proposed a deep attention U-Net (DAU-Net) model to automate the process of multi-organ segmentation for prostate cancer diagnosis via synthetic MRI, that is generated by processing the computed tomography scans using a cyclic generative adversarial network (CycleGAN)~\citep{zhu2017unpaired}. Initially, the CycleGAN model is trained to learn CT to MRI transformation which tends to add additional soft-tissue information without additional data acquisition techniques to produce sMRI data. Later, the sMRI data is used to train 3D DAU-Net model which incorporates conventional attention scheme~\citep{oktay2018attention} and deep supervision~\citep{wang2019deeply} with the U-Net model. The approach is trained and evaluated with 140 datasets from prostate patients to achieve $DSC$ value of 0.95, 0.87 and 0.89 for segmentation of bladder, prostate and rectum respectively, while also showing improvement over using raw CT images. 
	
	The prostate cancer diagnosis is another challenging task for which \cite{rundo2019use} proposed an automated approach, USE-Net, that uses the U-Net model by incorporating the squeeze-and-excitation (SE) blocks~\citep{hu2018squeeze} in skip connections to perform multi-class segmentation. Similar to the attention scheme~\citep{oktay2018attention}, the SE blocks tend to calibrate the channel-wise correlation while improving the generalization capability of the model across multi-institutional datasets. The USE-Net model outperformed its competitors when trained and evaluated on all datasets combined, where for other scenarios (individual dataset and mixed datasets), USE-Net struggled to achieve better results.  \cite{dong2020deu} integrated 3D U-Net model with deformable convolutions~\citep{dai2017deformable} for cardiac MRI segmentation. The deformable U-Net (DeU-Net) includes a temporal deformable aggregation module (TDAM) to generate fused feature maps using an offset prediction network. The fused feature maps are then fed to deformable global position attention (DGPA) network to map the multi-dimensional contextual information into generalized and localized features. The proposed approach outperformed other models to generate efficient segmentation masks involving subtle structures. Recently, \cite{li2021automatic} proposed multi-scale attention enabled U-Net model (MA-Unet) to segment lumbar spinar using MRI. The dual branch multi-scale attention block represents the most relevant feature maps at different target scales by using aggregating features obtained from two branches, where the first branch works as a multi-scale feature extraction block by using dense connections and the second branch uses the channel and spatial attention network to suppress irrelevant information. The authors achieved promising results; however, the introduction of MSA block across every stage in the entire network results in expensive computation and resources which could be reduced by using depthwise separable convolutions.

	\subsubsection{Inception U-Nets}
	\cite{chen2018s3d} improved the performance of the vanilla 3D U-Net model by adding spatiotemporal-separable 3D convolutions~\citep{xie2018rethinking} to form S3DU-Net model. The S3D convolution involves two convolution layers i.e. 2D convolution operation to extract spatial features and then additional 1D convolution to learn temporal features, furthermore, inception~\citep{szegedy2015going} and residual connections~\citep{he2016deep} are added to better learn the complex patterns. The S3DU-Net model is trained with dice loss and evaluated on dice coefficient and Hausdorff distance metrics. The authors achieved average dice scores of 0.69, 0.84 and 0.78, for enhancing tumor, whole tumor and tumor core respectively on BraTS 2018 challenge. For real-time applications, \cite{wang2019msu} proposed a multiscale statistical U-Net (MSU-Net) to segment cardiac regions in MRI. The MSU-Net incorporates statistical CNN (SCNN)~\citep{wang2019scnn} to fully exploit the temporal and contextual information present in various channels of an input image or feature map along with the multiscale parallelized data sampling approach. For multi-scale data sampling, independent component analysis (ICA)~\citep{wang2019scnn} is applied over the patches of data to form clusters of canonical form distributions which represent spatio-temporal correlations at coarser scales. This data sampling parallelization tends to speed up the performance significantly by 26.8\% as compared to the standard U-Net model and achieved an increased dice score by 1.6\% on ACDC MICCAI 2017 challenge, while also improving significantly over state-of-the-art GridNet~\citep{zotti2018convolutional} model.
	
	\subsubsection{Ensemble U-Nets}
	\cite{wang2019deeply} proposed a 3D FCN model with deep supervision and group dilation (DSD-FCN model) to address various challenges concerning the automated MRI prostate segmentation like inhomogeneous intensity distribution, varying prostate anatomy, etc., which makes it hard for manual intervention. The proposed architecture follows vanilla U-Net topology in which deep supervision is adopted to learn discriminative features, whereas group dilated convolutions tend to acquire multi-scale contextual information. The model is trained with the objective function defined as the weighted average of cosine similarity and cross entropy using the manually annotated institutional dataset and MICCAI PROMISE12 dataset, where authors achieved the $DSC$ values of 0.86 and 0.88 respectively. However, this achievement comes at the cost of complex computations due to group dilated convolutions \citep{wang2018understanding}. Recently, \cite{punn2020multi} proposed a 3D U-Net based framework for volumetric brain tumor segmentation. The proposed architecture is divided into three components: 1) multi-modalities fusion - to merge the MRI sequences with hierarchical inception convolution blocks, 2) tumor extractor - to learn the tumor patterns with 3D inception U-Net model using fused modalities, and 3) tumor segmenter - to decode the multi-scale extracted features into multi-class tumor regions. To achieve a better understanding of the input feature maps, each inception block aggregates multi-scale feature representation by using multiple filters equipped with short skip connections. With such dedicated components trained using a weighted average of dice and IoU loss functions, the authors achieved significant improvement over the existing approaches for BraTS 2017 and BraTS 2018 datasets. The increased computations due to inception convolution in each module could be reduced by using depthwise separable convolutions \citep{chollet2017xception}.
	
	\subsection{Positron emission tomography}
	The positron emission tomography~\citep{ollinger1997positron} is a widely used imaging in various clinical applications like oncology, brain, heart, etc., that helps in visualizing the biochemical and physiological reaction processes within the human body. The PET images are obtained by injecting a full dose of radioactive tracer or inhalation of gas to meet the clinical requirements. However, for minimal harm to human health, low-dose PET imaging is adopted to produce high quality imaging~\citep{wang20183d}.
	
	\subsubsection{Better U-Nets}
	With the huge success of U-Net in biomedical image segmentation, \cite{blanc2018automatic} demonstrated the potential of 3D U-Net model in $^{18}$F-fluoro-ethyl-tyrosine ($^{18}$F-FET) PET lesion detection and segmentation. F-FET PET/CT scans were acquired using a dynamic protocol from 37 patients, where the ground-truth segmentation masks were generated using manual delineation and binary thresholding. The 3D U-Net model comprises three stages of encoder and decoder paths with standard convolutions and pooling operations. The authors achieved a $DSC$ value of 0.79 on training and validation sets. However, the results could further be improved by increasing the data size with GAN based data augmentation techniques~\citep{frid2018gan} and other U-Net based approaches. Recently, \cite{lu2020automatic} proposed U-Net based automatic tumor segmentation approach in PET scans. The authors employed a transfer learning approach, where pre-trained VGG-19 blocks are added in the encoder phase to address the challenge of limited data availability. The authors adopted the DropBlock as a replacement for dropout to effectively regularize the convolution blocks. The model is fine-tuned using the Jaccard distance (IoU) as the loss function and the performance is validated with 1,309 PET images provided by the Shanghai Xinhua hospital (XH), which displayed improvements over the vanilla U-Net model.
	
	\subsubsection{Attention U-Nets}
	The integration of PET and CT modalities offer metabolic and anatomical information simultaneously. High contrast in PET scan enables the network to effectively segregate soft tissues around the tumor boundaries that are identified using CT imaging with high spatial resolution. In this context, \cite{fu2021multimodal} proposed multi-modal spatial attention module (MSAM) to segment tumor using PET-CT modalities. The MSAM block can be integrated with any U-Net model to leverage the multi-modalities features. Two individual encoder-decoder models are used, where one model is trained with PET imaging to generate spatial attention maps and another model utilizes this attention map at different scales by aggregating with each decoder layer to perform tumor segmentation using CT imaging. These PET based multi-scale attention maps guide the CT model towards the areas with high tumor likelihood. This approach of cross-modality multi-scale attention achieves promising results; however, the performance could further be improved by performing 3D volumetric analysis. 
	
	\subsubsection{Inception U-Nets}
	\cite{zhao2018tumor} proposed to utilize the multi-modalities (PET and CT) for computer-aided cancer diagnosis and treatment with the help of 3D FCN based V-Net~\citep{milletari2016v} model, which is an extension of the U-Net model for volumetric segmentation. A feature or intermediate level fusion approach is adopted, where two independent sub-segmentation networks are constructed to extract dedicated feature maps from each modality and are later fused with the cascaded convolution blocks that follow the V-Net model scheme to finally compute the tumor segmentation mask. The proposed framework is trained and validated on a limited clinical dataset of 84 patients suffering from lung cancer that consists of PET and CT imaging, where a dice value of 0.85 is achieved while outperforming other traditional models that use unary modality. In a similar approach, \cite{guo2019gross} adopted the fusion of PET and CT modalities to segment head and neck cancer (HNC) labelled as gross tumor volume (GTV). Due to resources limitations, the images are cropped during pre-processing. The authors utilized the modified 3D U-Net model in which the convolution blocks in encoder and decoder paths are replaced by dense convolution blocks~\citep{huang2017densely} that aggregates the multi-scale feature maps across various levels. The authors trained and evaluated the model on TCIA-HNC dataset, while achieving the $DSC$ value of 0.73 on the dedicated test set. The performance could be improved by analyzing complete 3D volume and obtain better affinities. 
	
	\subsubsection{Ensemble U-Nets}
	To address the need for a reliable and robust PET based tumor segmentation model, \cite{leung2020physics} proposed a novel physics guided deep learning based framework comprising three dedicated modules that segment each slice of PET volume to generate a complete mask. The first module tends to extract the realistic tumors with the available ground-truth boundaries via stochastic kernel-density estimation and physics based approach to generate simulated images. These images are fed to the improved U-Net model in the second module, which has minimal convolution and pooling blocks accompanied by dropout layers to aid in learning the complex features and generate efficient masks. Later, in the third module, the network is fine-tuned with delineation provided by the radiologist as surrogate masks to improve the learned features. The proposed framework achieved dice scores of 0.87 and 0.73 to segment primary tumors on simulated and patient images and outperformed several semi-automated approaches. This approach could perform 3D segmentation by generating a mask for each slice, but avoids the 3D correlations of the voxels which is crucial for real-time applications.
	
	\subsection{Ultrasound}
	Ultrasound is acoustic energy in the form of waves having a frequency beyond the human hearing range. These are generated with the help of piezoelectric crystals which deform under the influence of electric field and generate compression waves when an alternating voltage is applied. Ultrasonography~\citep{moore2011point} is an ultrasound based diagnostic imaging technique used for visualizing the internal body organs by processing the reflected signals. The deep learning technologies aid in diagnosing US imaging to segments regions of interest like breast mass, pelvic floor levator muscle discontinuity, etc. 
	
	\subsubsection{Better U-Nets}
	In consideration of breast cancer being the deadliest cancer among women, \cite{almajalid2018development} proposed an automatic breast ultrasound (BUS) image segmentation system to aid in its diagnosis and treatment. The authors utilized the vanilla U-Net model on the preprocessed BUS images. The images are preprocessed using the contrast enhancement with histogram equalization and noise reduction with speckle reducing anisotropic diffusion (SRAD)~\citep{yu2002speckle} techniques to improve the image quality. Finally, with the assumption of the presence of a single tumor region the authors filtered the false positive regions to remove the noisy regions. This assumptions limits the capability of the model to generate masks for multiple tumor regions. In this regard, \cite{li2019automatic} incorporated dense connections in the U-Net model (DenseU-Net) to efficiently segment levator hiatus from ultrasound images. The implication of dense connections enabled feature reuse and reduction in the trainable parameters. The DenseU-Net model is trained to generate the binary segmentation mask which is post-processed with binary thresholding to generate mask sharp boundaries, and localized regions are generated with active contour model~\citep{li2016segmentation} without compromising the performance of the model.
	
	\subsubsection{Attention U-Nets}
	In another application of eyeball segmentation, \cite{lin2019semantic} proposed a semantic embedding and shape-aware U-Net model (SSU-Net), where the authors employed a signed distance field (SDF) instead of a binary mask as the label to learn the shape information. In addition, the model is equipped with a semantic embedding module (SEM) to fuse the semantic information at coarser levels of the SSU-Net model. The SEM block draws features from two low-level stages and one corresponding stage, where lower level features are convolved and bilinear interpolation is applied to restore the resolution at the same scale. This enabled the network to efficiently identify the ambiguous and discontinuous boundaries and achieved better segmentation performance. Due to the low signal to noise ratio (SNR) in US imaging, real-time analysis is still a challenging task. Recently, \cite{zhang2020multi} proposed a U-Net based deep learning approach to realize the multi-needle segmentation in the 3D transrectal US (TRUS) images of high dose rate (HDR) prostate brachytherapy. The U-Net model is loaded with the attention scheme in the skip connections to address the challenge of identifying the smaller needles, while spatial continuity of the needles is maintained with total variation regularization. The model is trained with a deep supervision approach, where patches of needle masks are generated to compute the cross entropy loss and accordingly optimize the training weights. With the proposed framework, the authors achieved adequate performance gain on multi-needle segmentation for prostate brachytherapy.
	
	\cite{BYRA2020102027} proposed a selective kernel U-Net (SKU-Net) model for breast mass segmentation in US imaging while also addressing the challenge of variable breast mass size and image properties. In SKU-Net, each convolution layer of the U-Net model is replaced by an SK block, that tends to dynamically adapt the receptive field. Similar to the concept of dual path U-Net~\citep{yang2019robust}, the SK module~\citep{Li_2019_CVPR} is designed using two branches, where one uses dilated convolutions and other is without dilation to generate feature maps. Later, these features are merged and global average pooling, followed by FC layer and sigmoid activation is applied to construct attention coefficients for each channel in the feature map. With this approach, authors achieved significant improvement over the vanilla U-Net model across multiple datasets. In another work, \cite{punn2021rca} proposed residual cross-spatial attention (CSA) block in the skip connections of inception U-Net model \citep{punn2020inception} to further improve the segmentation performance. The authors validated the performance of the model with breast cancer segmentation using ultrasound imaging. In contrast to standard attention \citep{oktay2018attention}, the CSA block uses multi-level encoded feature maps to obtain better spatial correlation and develop spatial attention feature maps that are concatenated with corresponding decoder block to reconstruct the multi-scale spatial information. To address computational and memory challenges model uses depth-wise separable convolutions. The model achieved promising results on multiple US datasets and establishes scope for further utilization of this model across different modalities.
	
	\subsubsection{Inception U-Nets}
	In another approach, \cite{yang2019robust} proposed a dual path U-Net model for segmentation of lumen and media-adventitia from the IntraVascular UltraSound (IVUS) scans to aid in cardiovascular diseases diagnosis. Due to the limited availability of the data samples, the DPU-Net is trained with the real-time augmentor that generates and integrates three types of artefacts: bifurcation, side vessel, and shadow, and other common augmentation operations with training images. In contrast to vanilla U-Net, DPU-Net involves multi-branch parallel encoding and decoding operations, where feature maps are extracted and reconstructed with different kernel sizes at the same hierarchical level to address the challenge of a large variation in shape and size of lumen or media region. With this network-in-network architecture and real-time augmentation approach, the authors achieved Jaccard measure (IoU) of 0.87 and 0.90 on 40 MHz and 20 MHz frames respectively from IVUS dataset.  
	
	\subsubsection{Ensemble U-Nets}
	\cite{wang2018simultaneous} proposed a multi-feature guided CNN model for classification and segmentation of the bone surfaces in the US scans. The US images are initially processed with a pre-enhancing (PE) net to synthesize a US scan that highlights the bone surface, by using a B-mode US scan and three filtered image features, including local phase tensor image (LPT), local phase bone image (LB) and bone shadow enhanced image (BPE). The feature enriched images are then used by a classification embedded U-Net model (cU-Net) to produce the segmentation mask and identify the type of the bone surface. This multi-task deep learning framework achieved promising segmentation and classification results with F1-score of 0.96 and 0.90 on SonixTouch and Clarius C3 datasets respectively. In another application area, \cite{dunnhofer2020siam} emphasized on the tracking of knee femoral condyle cartilage during ultrasound guided invasive procedures. The Siam-U-Net model combines the potential of the U-Net model with siamese framework~\citep{gomariz2019siamese} for tracking the cartilage in the real-time ultrasound sequences. In Siam-U-Net two encoder blocks are adopted which are fed with resized-cropped US sequences named as, searching area and target cartilage. After five blocks of encoding layers, the acquired feature maps of two inputs are cross-correlated using convolution operation applied to searching area feature maps with target embedding as a filter, which results in localizing the implicit position of the cartilage in the searching area slice. Later, the slice is reconstructed in the decoder phase to generate the segmentation mask of the cartilage. The Siam-U-Net model achieved an average dice score of 0.70 with significant improvement over other approaches. However, the results could further be improved by expanding the dimension of the model into 3D space for considering the neighbouring voxels correlation.

	\section{Other U-Net variants and imaging}
	\label{sec4}
	In this section, various U-Net variants are presented that are introduced as the biomedical image segmentation networks, where each model acts as a generic architecture that is trained and evaluated on multiple/different modalities.
	
	\subsubsection{Better U-Nets}
	In the growing phase of biomedical image segmentation, \cite{alom2018recurrent} integrated the potential of multiple state-of-the-art deep learning models such as recurrent CNN~\citep{mikolov2011extensions}, residual CNN~\citep{he2016deep} and U-Net to form RU-Net and R2U-Net for BIS. In the RU-Net model, the standard convolution and up-convolution units are improved by incorporating recurrent convolutional layers (RCL), whereas in R2U-Net both RCL and residual units are added. These models are trained and evaluated on three different modalities such as retina blood vessel segmentation (DRIVE, STARE, and CHASH-DB1 datasets), skin cancer segmentation (ISIC 2017 Challenge), and lung segmentation (KDSB 2017 challenge). \cite{zhou2018unet} proposed a nested U-Net architecture, U-Net++, to narrow down the gap between the encoded and decoded feature maps. In contrast to the U-Net model, U-Net++ model follows convolutions on dense and nested skip connections to effectively capture the coarser details. Furthermore, a deep supervision approach is adopted to prune the model based on the loss (combined binary cross entropy and dice coefficient) estimated at different semantic levels. The performance of the model is validated with multiple datasets involving KDSB18, ASU-Mayo, MICCAI 2018 LiTS Challenge and LIDC-IDRI, while outperforming other models. \cite{azad2019bi} proposed another extension of U-Net, where bi-directional ConvLSTM (BConvLSTM) with densely connected convolutions (BCDU-Net) is introduced for BIS. The skip connections are equipped with BConvLSTM~\citep{song2018pyramid} to concatenate the feature maps between the encoded layer and the corresponding decoded layer. Furthermore, the dense connections are added at the bottleneck layer to extract and propagate features with minimal parameters. The authors achieved promising results across DRIVE, ISIC 2018 and LUNA datasets.	
	
	\subsubsection{Inception U-Nets}
	\cite{gu2019net} addressed the loss of spatial information while using the strided convolutions and pooling in U-Net with context-encoder network (CE-Net) to capture and preserve the information flow for BIS. In CE-Net the encoder unit is loaded with pre-trained ResNet blocks, the bottleneck layer (context extractor) includes dense atrous convolution (DAC) and residual multi-kernel pooling (RMP) blocks, and decoder block follows consecutive convolution and deconvolution blocks. The DAC module combines the design of Inception-ResNet-V2 model and atrous or dilated convolution, whereas RMP generates stacked feature maps followed from the pooling operations with varying window sizes to effectively model the target feature representations. This arrangement of operations achieved promising results on multiple modalities.
	
	For histopathological image segmentation, \cite{punn2020inception} proposed an inception U-Net model where standard convolution layers are replaced by inception blocks that consist of parallel convolutions of varying filter sizes and a hybrid pooling operation. The hybrid pooling operation draws the potential feature maps from the spectral domain via Hartley transform~\citep{zhang2018hartley} to preserve more spatial information and spatial domain with the help of max pooling to aim for sharp features, by using the $1\times 1$ convolution. The authors achieved significant improvement over other models using KDSB18 dataset with less number of parameters. \cite{ibtehaz2020multiresunet} proposed another extension of the U-Net model as MultiResU-Net, where the convolution operations are replaced with MultiRes blocks in encoder-decoder paths, and Res path is added in the bottleneck layer. Inspired from the inception and residual model, the MultiRes blocks are built using stacked convolutions with a succession of $3\times 3$ filters, and a residual $1\times 1$ convolution connection is added. The Res path tends to propagate the feature maps from the encoder phase to decoder phase with the series of residual convolution blocks. The model is evaluated on different datasets covering fluorescence images, ISBI-2012, ISIC-2017, CVC-ClinicDB and BraTS17.
	
	\subsubsection{Attention U-Nets}
	In most of the U-Net models, the long-range dependencies are gradually acquired with local convolutions which limit the overall efficiency and effectiveness of the model. Inspired from the transformer models \citep{vaswani2017attention}, \cite{wang2020non} proposed non-local Unet (NL-Unet) that comprises of self attention based global context aggregation module to extract full context information which can be easily integrated with feature extraction and reconstruction operation in any U-Net model. The traditional spatial \citep{oktay2018attention} and channel attention \citep{hu2018squeeze} mechanisms lack to establish correlations with different targets and features, due to which non-local attention based models exhibit potential to perform better in biomedical image segmentation.
	
	\subsubsection{Ensemble U-Nets}
	With the immense application of U-Net model in the medical domain, \cite{isensee2021nnu} proposed a self-adapting framework, no-newU-Net (nnU-Net) to establish the generalized architecture and training mechanism for vivid modalities, inspired by the medical segmentation decathlon (MSD) challenge. The nnU-Net framework comprises an ensemble of 2D U-Net, 3D U-Net and 3D U-Net cascade, along with an automated pipeline to adapt the requirements of the dataset such as preprocessing, data augmentation and post-processing. The model achieved state-of-the-art segmentation results without manual intervention for different modalities in the medical segmentation decathlon challenge.
	
	With the recent success of transformer models in sequence-to-sequence modelling~\citep{vaswani2017attention}, it has also been integrated with U-Net for medical image segmentation and has achieved satisfactory results. In vanilla U-Net, there is limited learning of global context information due to the local convolutions and hence, it cannot capture long-range dependencies. To address this challenge, \cite{cao2021swin} proposed Swin-Unet which is a Unet-like pure transformer architecture. This model uses Swin transformer \citep{liu2021swin} with shifted windows as the encoder for feature extraction and patch-expanding Swin transformer for restoration of image resolution. In similar context, to further improve the performance, \cite{wang2021mixed} introduced a mixed transformer module (MTM) in U-Net that refined the self attention mechanism by simultaneously obtaining intra- and inter-correlations while using local-global Gaussian-weighted self attention (LGG-SA) and external attention, respectively. These MTM blocks are arranged in U-Net topology for medical image segmentation. Though these models achieved better results; however, rely on large-scale pre-training and have high computational complexity. 
	
	\section{U-Net in COVID-19 diagnosis}
	\label{sec5}
	The ongoing pandemic of the severe acute respiratory syndrome - coronavirus (SARS-CoV-2) also known as COVID-19 has brought the worldwide crisis along with the rampant loss of lives. This contagious virus initiated from Wuhan, the People’s Republic of China in December 2019 and till November 17, 2021, have caused 254,174,536 infections and 5,112,325 deaths worldwide~\citep{JHU}. Currently, the most reliable COVID-19 diagnosis approach follows the reverse-transcriptase polymerase chain reaction (RT-PCR) testing; however, it is time consuming and less sensitive to identify the virus at the early stages. 
	
	With the advancements in the technology and data acquisition systems~\citep{agarwal2020unleashing, shi2020review}, deep learning based approaches are developed to assist in the COVID-19 diagnosis with the help of CT and X-ray modalities~\citep{COVID19} to control the exponential growing trend~\citep{punn2020covid} of the spread. \cite{wu2020jcs} proposed a JCS framework (similar to cU-Net) for joint classification and segmentation of COVID-19 from chest CT scans using the U-Net model. In another U-Net based implementation, a feature variation block is introduced in the COVID-SegNet model~\citep{yan2020covid} to better segment the COVID-19 infected regions by highlighting the boundaries and diverse infected regions. The lung infection segmentation deep network (Inf-Net)~\citep{fan2020inf} followed U-Net topology with diverse modifications including reverse attention and parallel partial decoder. The authors validated the performance in the supervised and semi-supervised modes to address the challenge of limited availability of the labelled data. Recently, \cite{punn2020chs} introduced a hierarchical segmentation approach, CHS-Net that involves two cascaded residual attention inception U-Net (RAIU-Net) models, where first generates lungs contour, which is fed to the second model to identify COVID-19 infected regions using CT images. The RAIU-Net model is designed with a residual inception U-Net model and spectral-spatial-depth attention blocks. The authors achieved promising results in generating the infected segmentation masks. 
	
	Furthermore, similar approaches are also developed for X-ray imaging for the screening of COVID-19~\citep{punn2020automatedd}. \cite{zahangir2020covid_mtnet} proposed a robust classification and segmentation framework of coronavirus infected X-ray and CT images, where classification is performed using inception residual recurrent convolutional neural network (IRRCNN)  with transfer learning and NABLA-N model is used for localizing the infected regions. In addition, other deep learning based application areas are also explored to control the spread of the virus such as automated social distancing monitoring~\citep{punn2020monitoring}, mask detection~\citep{chowdary2020face}, etc. Furthermore, the survey of deep learning based approaches for COVID-19 diagnosis~\citep{shi2020review} reveals the significant impact of U-Net for CAD systems. Following these developments, it is believed that these artificial intelligent approaches will continue to evolve and contribute towards the faster and efficient diagnosis of the coronavirus. 
	
	\section{Analysis}
	\label{sec6}
	Over the years, the advancements in deep learning and computer vision techniques have attracted many researchers to contribute to the healthcare domain with a variety of tasks e.g. classification, detection, segmentation, etc. With segmentation being a critical task that drives the diagnosis process~\citep{hesamian2019deep}, researchers have developed a keen interest to develop a computer-aided diagnosis system to speed up the treatment process.
	
	\begin{table}
		\centering
		\caption{Summary of popular BIS datasets.}
		\label{tab7}
		\resizebox{\columnwidth}{!}{\begin{tabular}{p{0.6in}p{2in}p{1.7in}}
				\hline
				\textbf{Dataset}                 & \textbf{Description}                                                                           & \textbf{Availability}                                                 \\
				\hline
				ISBI 2012                        & Electron microscopy   cell slides for cell segmentation                                        & \url{http://brainiac2.mit.edu/isbi\_challenge/}                             \\
				CVC-ClinicDB & Endoscopic colonoscopy frames for polyp detection & \url{https://www.kaggle.com/balraj98/cvcclinicdb} \\
				ISBI                             & 2D and 3D videos of moving   cells for cell tracking                                           & \url{http://celltrackingchallenge.net/}                                     \\
				KDSB 2018                        & Histopathological cell images   for nuclei segmentation                                        & \url{https://www.kaggle.com/c/data-science-bowl-2018}                       \\
				PanNuke                          & Histopathological slides for   nuclei segmentation                                             & \url{https://jgamper.github.io/PanNukeDataset/}                             \\
				DRIVE                            & Retinal fundus images for   vessel extraction                                                  & \url{https://drive.grand-challenge.org/}                                    \\
				STARE                            & Retinal fundus imaging for blood vessel segmentation                                  & \url{http://cecas.clemson.edu/\%7Eahoover/stare/}                           \\
				CHASE\_DB1                       & Retinal fundus imaging for blood vessel segmentation                                  & \url{https://blogs.kingston.ac.uk/retinal/chasedb1/}                        \\
				LiTS                             & Liver CT scans for tumor segmentation                                                          & \url{https://competitions.codalab.org/competitions/17094}                   \\
				LIDC-IDRI                        & Lung CT scans for cancer segmentation                                                          & \url{https://wiki.cancerimagingarchive.net/display/Public/LIDC-IDRI}        \\
				LUNA 2016                        & CT scans for lung nodule   segmentation                                                        & \url{https://luna16.grand-challenge.org/}                                   \\
				xVertSeg                         & CT spine images for vertebra   segmentation                                                    & \url{http://lit.fe.uni-lj.si/xVertSeg/}                                     \\
				SIIM-ACR                         & Chest X-rays for pneumothorax   segmentation                                                   & \url{https://www.kaggle.com/c/siim-acr-pneumothorax-segmentation/data}      \\
				ISIC                             & Dermoscopy images for skin   lesion segmentation                                               & \url{https://www.isic-archive.com/}                                         \\
				BraTS  2012 – 2020               & MRI modalities (T1, T2, FLAIR)   for brain tumor segmentation.                                 & \url{http://braintumorsegmentation.org/}                                    \\
				ISLES                            & MRI scans for stroke lesion   segmentation                                                     & \url{http://www.isles-challenge.org/}                                       \\
				ICCVB                            & Prostate MRI and retinal   fundus imaging                                                      & \url{http://i2cvb.github.io/}                                               \\
				IBSR                             & Repository of MRI imaging                                                                      & \url{https://www.nitrc.org/projects/ibsr}                                   \\
				ACDC 2017                        & MRI imaging for cardiac   diagnosis and segmentation                                           & \url{https://www.creatis.insa-lyon.fr/Challenge/acdc/index.html}            \\
				PROMIS 2012                      & Prostate MRI image   segmentation                                                              & \url{https://promise12.grand-challenge.org/}                                \\
				Med. Seg. Decathlon & MRI and CT modalities for tumor   segmentation in various organs like liver, brain, lung, etc. & \url{http://medicaldecathlon.com/}                                          \\
				OASIS                            & MRI and PET images for aging   analysis and segmentation                                       & \url{https://www.oasis-brains.org/}                                         \\
				Head-Neck-PET-CT                 & PET and CT imaging for tumor   segmentation                                                    & \url{https://wiki.cancerimagingarchive.net/display/Public/Head-Neck-PET-CT} \\
				BUSIS                            & Ultrasound imaging for breast   tumor segmentation                                             & \url{http://cvprip.cs.usu.edu/busbench/}                                    \\
				BUSI                             & Breast ultrasound scans for tumor segmentation                                             & \url{https://scholar.cu.edu.eg/?q=afahmy/pages/dataset}\\
				\hline                    
		\end{tabular}}
	\end{table}
	
	Among the published approaches or frameworks, U-Net appears to be the prominent choice~\citep{minaee2020image} to develop novel architectures to adapt multiple modalities with optimal segmentation performance. Following such high utility of the model, this article presented the recent developments in U-Net based approaches for biomedical image segmentation. Due to the high mutability and modularity design, U-Net topology can easily be integrated with other state-of-the-art deep learning models such as AlexNet~\citep{krizhevsky2012imagenet}, VGGNet~\citep{simonyan2014very}, ResNet~\citep{he2016deep}, GoogLeNet~\citep{szegedy2015going}, MobileNet~\citep{howard2017mobilenets}, DenseNet~\citep{huang2017densely}, etc., to produce the desired results depending on the application. This ease of integration opens a wide spectrum of applications for U-Net with endless possibilities of novel architecture designs. In the most recent developments of U-Net based biomedical image segmentation models following observations are drawn: 
	\begin{itemize}
		\item 
		More emphasis is given to multi-scale feature extraction and fusion to explicitly model global and long-range feature dependencies.
		\item
		Inspired by the state-of-the-art performance of self attention mechanism in transformer models many transformer based U-Net variants are utilized to enhance its capability to capture global contexts.
		\item
		For the training phase, most models employed a hybrid loss function that combines the binary cross entropy loss with dice similarity coefficient loss or with Jaccard loss, which tends to better penalize the false positive and false negative predictions
		\item
		Considering the implementation strategies mostly authors applied an end-to-end training-from-scratch approach with minimal pre-processing i.e. resizing and normalization and without any post-processing.
		\item
		Mostly depthwise separable convolutions are employed to reduce the overall number of computations and training parameters of the model.
		\item
		Multi-modality fusion based approaches are also developed for better feature representation learning concerning target regions.
		
	\end{itemize}
	
	From the reviewed articles it is observed that some of the segmentation approaches utilize the local dataset (datasets that are not publicly accessible), which tend to limit their reusability and reachability. In order to develop a widely acceptable solution, the summary of most widely utilized publicly available datasets for BIS is provided in Table~\ref{tab7}. These benchmark datasets aid the research community to validate the existing performance and propose further improvements. Among the reviewed articles, CT and MRI modalities cover a wide range of U-Net variants for biomedical image segmentation. Moreover, for PET scan and ultrasound imaging most of the proposed approaches are validated on the local dataset, whereas for X-rays the approaches aim to localize the target structure with the bounding boxes. Despite such variants, it is difficult to conduct an effective comparative analysis of the results because each approach is evaluated with different evaluation metrics such as accuracy, F1-score, Jaccard index, etc. However, among these metrics, dice similarity coefficient is most widely utilized to quantize segmentation performance.
	
	Considering the present survey it is also observed that each modality requires a different approach to address the corresponding challenges. Though there are segmentation approaches that are validated on multiple modalities to form generic architectures like nn-UNet, U-Net++, MR-Unet, etc. but it is difficult to achieve optimal performance in all segmentation tasks. The main reason is due to the diverse variation in the features corresponding to the target structures involving lungs nodule, brain tumor, skin lesions, retina blood vessels, nuclei cells, etc. and hence require different mechanisms (dense, residual, inception, attention, fusion, etc.) to integrate with U-Net model to effectively learn the complex target patterns. Moreover, the presence of noise or artefacts in different modalities adds another factor to propose different segmentation methods.
	
	\begin{figure}
		\centering
		\includegraphics[width=\columnwidth] {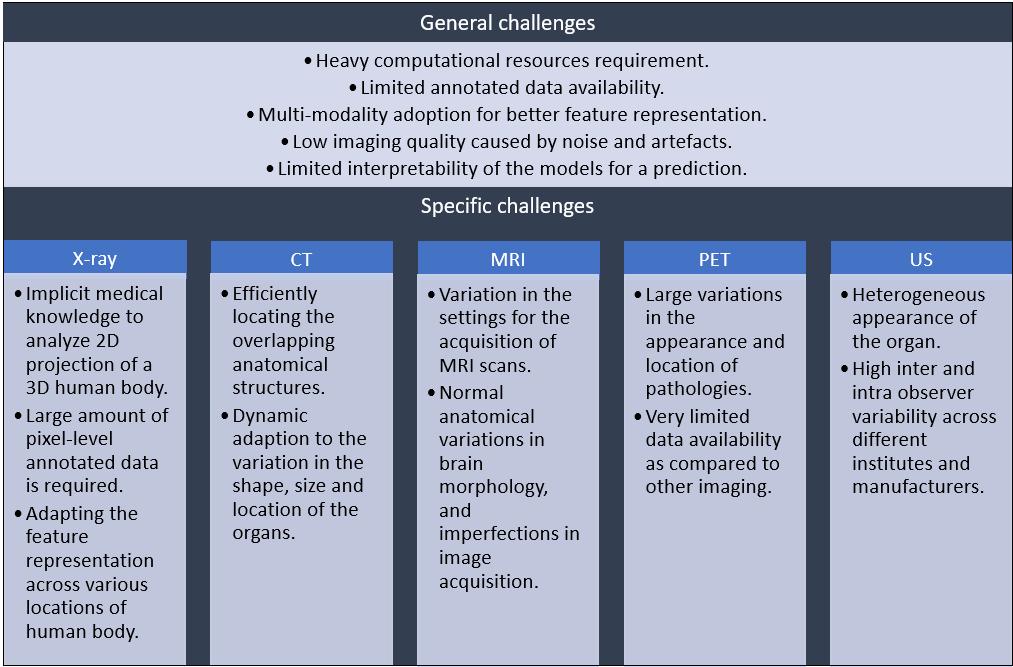}
		\caption{Challenges involved in biomedical image segmentation.}
		\label{fig8}
	\end{figure} 
	
	\section{Scope and challenges}
	\label{sec7}
	Deep learning technologies have played a vital role in advancements towards medical diagnosis and applications. Generally, deep learning based technologies such as U-Net aims to develop CAD systems to achieve the desired results with minimal error. Despite U-Net based models being superefficient for biomedical image segmentation, there are various challenges involved, as shown in Fig. \ref{fig8} with general and modality specific challenges, for developing the real-world implications of the deep learning models. With regular advancements in deep learning, these challenges are tackled with hardware and software oriented approaches which consequently attracts researchers to develop novel architectures and frameworks for biomedical image segmentation.
	
	\subsection{General challenges}
	One major challenge is concerned with the computational power requirement which tends to limit the feasibility of the approach. Following this many cloud based high performance computing environments are developed for mobile, efficient and faster computations. Although progress is also made towards the model compression and acceleration techniques~\citep{cheng2017survey} with great achievements; however, it is still required to establish the concrete benchmark results for real-time applications. Recently, \cite{tan2019efficientnet} proposed an EfficientNet framework that uses compound coefficients for uniform scaling in all dimensions. This could make the U-Net design streamlined for complex segmentation tasks with minimal change in the parameters. Besides several attempts are also made towards automation of model
	architecture design \citep{ren2021comprehensive} to develop optimal model for different applications; however, there is still long way to go.
	
	Furthermore, these powerful deep learning approaches are data-hungry i.e. the amount of data available directly affects the model performance towards achieving robust results. However, the expense of data acquisition and delineation, and data security, results in the limited availability of the data which bottlenecks the development of real-world systems. In this context, various data augmentation strategies~\citep{shorten2019survey} are proposed that tend to alleviate the performance of the model while drawing the advantages of big data. Generally, the image augmentation strategies involve geometric transformations, color space augmentations, kernel filters, mixing images, random erasing, feature space augmentation, adversarial training, generative adversarial networks, neural style transfer, and meta-learning. However, the diversity of augmented data is limited by the available data which could result in overfitting. In another approach, U-Net models utilize transfer learning approaches~\citep{byra2020knee} to optimize the pre-trained model to adapt to the targeted task while having insufficient training data. These deep transfer learning techniques are categories under four broad areas: instances based, mapping based, network based and adversarial based~\citep{tan2018survey}. The self-supervised learning (SSL)~\citep{jing2020self} is an emerging technology that also addresses this challenge. In SSL strategies initially, pre-training is performed with un-annotated samples for some pretext task to learn feature representations such as predicting rotations, identifying the image patch, solving jigsaw puzzles, etc. and later model is fine-tuned to perform actual segmentation. The potential of this approach attracts many researchers to advance the U-Net based BIS approaches. Furthermore, with the fusion of different modalities, rich information can be extracted concerning desired features for training the model. However, developing an appropriate fusion approach representing vivid modalities is still a challenging task.
	
	The performance of models are also affected by the low imaging quality caused by the noise and artefacts, where noise may obscure features of an image, while artefact adds irrelevant features following some pattern. For instance in CT imaging, noise can make images grainy with small variations in contrast, whereas a streak artefact appears to make the region of low density. There are several pre-processing strategies proposed to remove or minimize the presence of noise and artefacts from data. For denoising the most common approaches that are followed are wavelets thresholding, partial differential equations (PDEs) (minimization problem of total variation method), NL-means and fast NL-means algorithms, anisotropic diffusion, etc.~\citep{2012Oulhaj, ravishankar2017survey}. The artefacts can be reduced by using newer reconstruction or metal artefact reduction (MAR) techniques~\citep{chen2019analytical, triche2019recognizing}.   
	
	In general, the decision made in the rule-based applications can be traced back to its origin; however, deep CNN models lack transparency in the decision making process, where the input and output are well-presented but the processing in the hidden layers is difficult to interpret and understand, and hence these are also termed as black-box models. To better interpret these models various visualization based approaches are proposed such as local interpretable model-agnostic explanations (LIME)~\citep{mishra2017local}, shapley additive explanation (SHAP)~\citep{lundberg2017unified}, partial dependence plots (PDP)~\citep{friedman2001greedy}, anchor~\citep{ribeiro2018anchors}, etc. Currently, these approaches are applied to explain and interpret the obtained results from deep learning models, but still, a concrete benchmark scheme is required to be established.
	
	\subsection{Specific challenges}
	In addition to the general challenges each modality also exhibit unique challenges. Some of these major challenges that are profound in X-ray, CT, MRI, PET and US imaging are shown in Fig. \ref{fig8}. In X-ray imaging because of 2D projection of the 3D human body, features representing physiological structures overlap each other which may result in variation in the anatomy representation. For instance, in chest X-rays, due to the presence of scarring the lung contours are substantially blurred and hence segmentation models must learn the global concept for resolving the ambiguities and producing the correct mask. This challenge is mostly addressed by using U-Net based models that use long skip connection variants to transfer the knowledge from extraction to reconstruction phase~\citep{rashid2018fully, frid2018improving}, and adversarial learning strategies~\citep{gaal2020attention}. However, such model requires a large amount of pixel-level annotated training data which can be addressed by using data augmentation, self-supervised learning \citep{punn2021bt} or semi-supervised learning \citep{yu2018adaptive} strategies. In another challenge, a model needs to be designed with dynamic feature adaption to generate segmentation masks at different locations in the human body. It can be considered as one of the important aspects for bone segmentation~\citep{wang2020mdu} while using X-ray imaging.
	
	With CT imaging most of the challenges arise due to overlapping anatomical structures and large variations in the shape, size and location of the organs from person to person. For example, in the case of an abnormal lung CT segmentation, lung parenchyma~\citep{skourt2018lung} needs to be segregated from the bronchus regions that represent similar features as lung tissue, along with the segmentation of nodules and blood vessels. Moreover, pulmonary inflation with an elastic chest wall can result in large variation in volumes and margins~\citep{mansoor2015segmentation}. To address these challenges mostly attention based U-Net models are employed in CT image segmentation~\citep{fan2020ma, seo2019modified, song2019u}, while the performance could further be improved by incorporating other networks or operational designs such as object dependent filters, residual blocks, etc.
	
	Unlike other modalities, MRI is most widely used for segmentation (brain tissue, tumor, skull, prostate, etc.) due to the ample availability of datasets. However, automated analysis using MRI is challenging due to intensity inhomogeneity, changes in settings for the acquisition of MRI scans, fluctuations in the appearance of pathology, anatomical variations in brain morphology, and imperfections in image acquisition. For instance, the performance of brain tumor segmentation models is affected by large variations in brain tumors location, size, shape and heterogeneity (image uniformity, contrast uptake and texture)~\citep{akkus2017deep, wadhwa2019review}. To address these challenges multi-modality fusion based approaches are most widely studied to effectively learn the inconsistent tumor features~\citep{zhou2019review}. Depending on the segmentation problem under consideration these challenges can be addressed by integrating several state-of-the-art architectural designs (inception, cascaded, attention, dense, etc.) and operational designs (atrous, spectral, hybrid, etc.) resulting in ample possibilities of approaches~\citep{dong2020deu, punn2020multi, zhang2019automatic}. Similar to MRI, PET imaging analysis is mostly utilized by oncologists to diagnose and analyse severe problems such as gliomas, perfusion evaluation, cerebrovascular accidents, parkinson's disease, etc. This brings similar challenges of large variations in the appearance and location of pathologies~\citep{weller2013molecular} in addition to very limited data availability due to privacy and security concerns which can be addressed by using U-Net based models assisted with data augmentation, self-supervised learning or transfer learning strategies~\citep{lu2020automatic}. 
	
	There are various clinical applications of ultrasound imaging including cardiology, breast cancer, prostate, and other diseases~\citep{noble2006ultrasound} which can be assisted with automated segmentation. However, the heterogeneous appearance of the organ due to variations in depth, neighbouring tissues and location is one of the major challenges in ultrasound image segmentation~\citep{zhou2020artificial}. In this regard, most of U-Net based approaches uses inception based attention with dense encoder modules to develop multi-scale feature representation~\citep{yang2019robust, li2019automatic, wang2018simultaneous}. Another major challenge is concerned with the high variability in the inter and intra-observer among physicians and sonographers, which depends on the acquisition protocols and observer preference, thereby a larger training dataset is required to alleviate the variation~\citep{liu2019deep}.

	\section{Conclusion}
	\label{sec8}
	The deep learning approaches especially U-Net has great potential to influence the clinical applications involving automated biomedical imaging segmentation. With U-Net being a breakthrough development, it sets up the foundation for the development of novel architectures concerning the identification and localization of the target regions or sub-regions. Following from this context, in present article, various U-Net variants are explored, covering current advancements and developments in the area of biomedical image segmentation serving different modalities. Each U-Net variant features unique developments over the challenges incurred due to different modalities. With such high utility and potential of the U-Net models, it is believed that U-Net based models would be widely applied to address various challenging problems experienced in the biomedical image segmentation for developing real world computer-aided diagnosis systems.
	
	\section*{Acknowledgment}
	We thank our institute, Indian Institute of Information Technology Allahabad (IIITA), India and Big Data Analytics (BDA) lab for allocating the necessary resources to perform this research. We extend our thanks to our colleagues for their valuable guidance and suggestions.
	
	\bibliographystyle{spbasic}
	\bibliography{reference}
	%
	%


\end{document}